\documentclass{llncs}
\usepackage{hyperref}

\usepackage{graphicx}
\usepackage{url}
\usepackage{subfigure}
\usepackage{verbatim}

\usepackage{wrapfig}
\usepackage{xspace}
\usepackage{cite}
\usepackage{color}
\usepackage{textcomp} 
\long\def\omitthis#1{\relax}

\long\def\longversion#1{#1}
\long\def\shortversion#1{\relax}

\def\sep{\mathit{sep}}

\def\2sd{\mathit{2sd}}

\def\delta#1{\mathit{delta.#1}}

\title
{Transcriptional Response of \\ SK-N-AS Cells to Methamidophos \\
	Extended Version\thanks{Sponsored by the US Army Research Office and the Defense Advanced Research Projects Agency; accomplished under Cooperative Agreement W911NF-14-2-0020.}
}

\author
{
Akos Vertes\inst{1} \and 
Albert-Baskar Arul\inst{1} \and
Peter Avar\inst{1} \and
Andrew R. Korte\inst{1} \and 
Lida Parvin\inst{1} \and 
Ziad J. Sahab\inst{1} \and 
Deborah I. Bunin\inst{2} \and 
Merrill Knapp\inst{2} \and 
Denise Nishita\inst{2} \and 
Andrew Poggio\inst{2} \and 
Mark-Oliver Stehr\inst{2} \and 
Carolyn L. Talcott\inst{2} \and 
Brian M. Davis\inst{3} \and 
Christine A. Morton\inst{3} \and 
Christopher J. Sevinsky\inst{3} \and
Maria I. Zavodszky\inst{3}
}

\institute
{
Dept. of Chemistry, George Washington Univ., Washington, DC 20052\and 
SRI International, Menlo Park, CA 94025\and
GE Global Research, Niskayuna, NY 12309
}
 
\begin{document}

\maketitle


\begin{abstract}

Transcriptomics response of SK-N-AS cells to methamidophos
(an acetylcholine esterase inhibitor)
exposure was measured at 10 time points between 0.5 and 48 h.
The data was analyzed using a combination of traditional
statistical methods and novel machine learning algorithms 
for detecting anomalous behavior and infer causal relations 
between time profiles. We identified
several processes that appeared to be upregulated in cells treated
with methamidophos including: unfolded protein response, response to cAMP,
calcium ion response, and cell-cell signaling. The data confirmed the expected consequence of acetylcholine buildup. In addition,
transcripts with potentially key roles were identified and causal
networks relating these transcripts were inferred using two different
computational methods: Siamese convolutional networks and time warp causal inference. Two types of anomaly detection algorithms, one based on Autoencoders and the other one based on Generative Adversarial Networks (GANs), were applied to narrow down the set of relevant transcripts.

\end{abstract}

\section{Introduction}\label{intro}

Rapid determination of the mechanism of action (MoA) of an unknown or
novel xenobiotic (toxin, drug, pathogen) and its consequences is
important both scientifically and for biodefense. It is particularly
important to develop methods to identify candidates that are
independent of existing experimental data, as there may be no such
data available.

Time series data generated by omics experimental techniques provides a
wealth of data about change in relative concentrations of transcripts,
proteins and metabolites. For example, chemically perturbing cells can
result in thousands of mRNAs with at least a 2 fold expression change.
The challenge is to get the most information purely from the data,
before augmenting the conclusions with knowledge from databases and
literature. Thus, it is important to consider not only what changes,
but how it changes over time, to identify key responders and how they
organize into cellular processes.

As part of the DARPA Rapid Threat Assessment project we developed a
suite of data analysis methods to identify candidate biological
players and processes that make up the cellular response to a
challenge. These included traditional statistical analysis,
shape/feature analysis, Gaussian process representation, new machine
learning algorithms for identifying anomalies and inferring causal
relations between time profiles. The algorithms and methods were
developed using data from HepG2/C3A cells exposed to a series of
different drugs each affecting different known cellular processess. To
test the robustness and generality of the analysis methods we selected
a different cell type (SK-N-AS human neuroblastoma cells) and toxin
(the organophosphate methamidophos). We expected the biological noise
to be different in a different cell type. We also expected the timing
and organization of response to an organophosphate to be different
from the previously tested drugs. This allows checking that parameters
chosen for multiple algorithms work in a more general setting. Here we
present three analysis algorithms not previously described, and
discuss the application of our suite of algorithms to analysis of
methamidophos response data.

Methamidophos is a cholinesterase inhibitor. The enzymes
acetylcholinesterase (ACHE) and butyrylcholinesterase (BCHE)
convert acetylcholine into the inactive metabolites choline and
acetate. The result of acetylcholine esterase inhibition in
cultured cells is that acetylcholine builds up and continues to
bind and activate muscarinic and nicotinic acetylcholine receptors.

Transcriptomic response of SK-N-AS cells to methamidophos exposure was
measured at 10 time points between 0.5 and 48 h. The data was analyzed
using our suite of algorithms. This lead to multiple classifications
of transcripts and groups of transcripts as candidate elements of the
methamidophos broader~\footnote{By broader here we mean the processes
the cell is using to respspond to a detected challenge, going beyond
the initial entry or binding mechanism.}
MoA. GO process term annotations (using 
UniProt~\cite{uniprot-2019nar}
or PatherDB~\cite{mi-etal-2017panther}), databases and curated
experimental results were used to refine the data analysis and
determine possible biological roles of the identified transcripts.
Several biological processes were hypothesized as candidate components
of the overall mechanism of action, including unfolded protein
response, response to cAMP, and calcium ion related processes. The
data also shows response to an increase in the second messenger
diacylglycerol (DAG) which is consistent with acetylcholine build up.
However, the DAG response was not identified by over-expression
analysis, partly because of a lack of relevant GO annotations.
Transcripts with potentially key roles were identified and causal
networks relating responsive transcripts were inferred. Some of the
inferred causal edges correspond to known relations, most are novel
and more work is needed to understand what they mean.

\paragraph{Plan }
In Section \ref{data-analysis} we give an  overview of the
data analysis process and describe the novel classification
and causality inference algorithms.
The main results of the data analysis are discussed in
Section \ref{results}.  Details of the experiments generating
the data are given in Section \ref{materials-methods}.  Section \ref{discussion} contains
some discusion of results, related and future work.
The appendix \ref{appendix} provides some additional data
analysis details.

\section{Overview of Data Analysis}\label{data-analysis}

\begin{figure*}[h]
\centering 
\includegraphics[width=.9\linewidth]{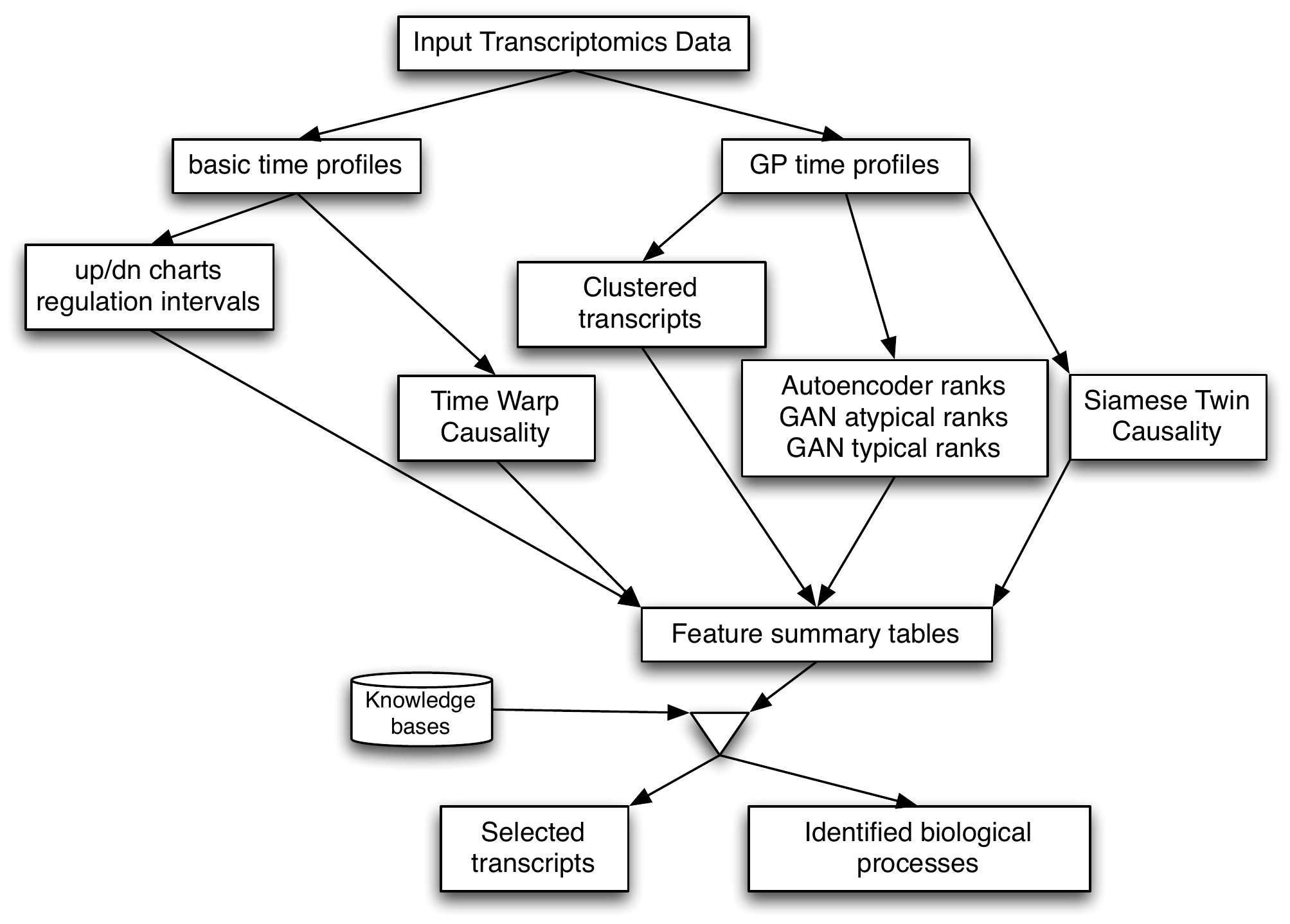}
\caption{Data analysis  schematic}
\label{workflow}
\end{figure*}

Figure~\ref{workflow} shows a schematic of our data analysis
process. The left branch uses log$_2$ fold change (basic) time profiles
derived from the means of the control and treated signals
in the usual way. Up/down charts map transcripts to the first
time point the log$_2$ fold change magnitude passes $1$.
Regulation intervals delimit times that log$_2$ fold change
stays above $0.75$ or below $-0.75$. The right branch uses time
profiles obtained by Gaussian process (GP) modeling
\cite{vertes-etal-2018cmsb}. Using these time profiles,
transcripts are clustered (k-means using PCA to reduce
dimensionality), and ranked by contribution to PCA components
and by two
machine learning algorithms, one using autoencoder techniques
(see \cite{vertes-etal-2018cmsb}) and one using Generative
Adversarial Nets (GANs, \cite{gans}, Section \ref{mm-gan}). 
Transcripts are ranked highly as
anomalies according to how poorly they are reconstructed from
the autoencoding, or how unsure the trained GAN
discriminator is that the time profile represents a transcript.
Transcripts are also given a `real/typical' ranking according
to how confident the GAN discriminator is that the time profile
represents a transcript. Two algorithms were used to infer potential `causal' edges between time profiles. The Siamese twin causality detection
algorithm~\cite{canes} is based on two Siamese neural
networks~\cite{Bromley94}. One Siamese network is trained to
detect undirected causality; the other is trained to detect
lag. Lag detection is used to direct the undirected causality
edges (see Section \ref{mm-twin}). 
The Timewarp algorithm operates on
basic time profiles and uses a variant of the Needleman-Wunsch
alignment algorithm~\cite{needleman-wunsch-1970jmb} to align
time  profiles (see Section \ref{mm-bayes}). 
The results of the analyses, along with an indication of
satisfaction of several significance filters, are collected in
a `feature' table that can be sorted to highlight features of interest.
We used the ranking functions and significant change filters to  select a set of transcripts, called Top20X, as
the starting point for further identification of MoA candidates.
We consider two categories: biological processes, individual transcripts.
To identify candidate processes we used
PatherDB over representation analysis~\cite{mi-etal-2017panther} combined with our
GO term annotations of k-means clusters~\cite{vertes-etal-2018cmsb}.
Potentially key transcripts were selected from 
the Top20X transcripts using a combination of ranking, clustering,
and GO annotations.

\subsection{The input data}

Transcriptomic response was measured at ten time points (0.5, 1,
2, 4, 6, 8, 18, 24, 32, and 48 h) after treatment of SK-N-AS cells
with and without methamidophos (2mg/ml) using 3 technical
replicates.~\footnote{We use the term \emph{technical replicate} as
defined by NIH: Technical replicates are repeated measurements of
the same sample that represent independent measures of the random
noise associated with protocols or equipment.} This resulted in a
data set with measurements for 3 treated samples and 3 control
samples for each time point for each transcript using microarray
based technology. The analysis software produces log$_2$ intensities
for more than 67,000 transcripts. We restricted
attention to the approximately 18,000 transcripts that are known to
code proteins.

\subsection{Gaussian Process Modeling}

\begin{figure*}[h]
	\vspace{-10pt}
	\centering 
\includegraphics[width=.9\linewidth]{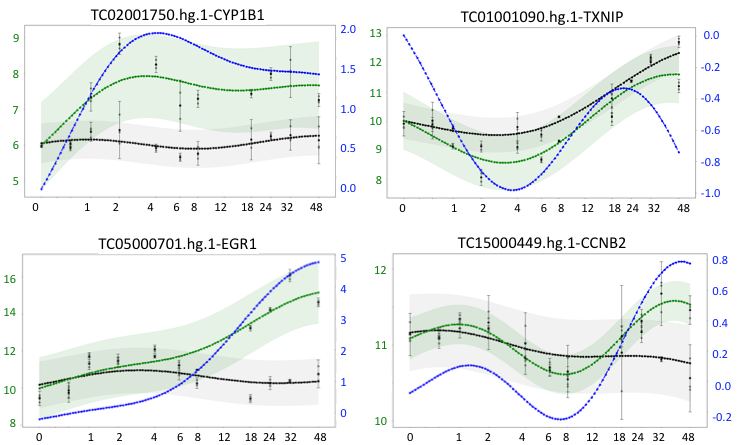}
\caption{GP profiles for CYP1B1, TXNIP, EGR1, and CCNB2. The dots with whiskers are the input data. The black/green dotted
line is the control/treated mean value (all log$_2$), sampled at
100 time points between 0 and 48 h. The blue dotted line is the
(log$_2$) ratio (the difference between the green and black). The
grey/green bands around the control/treated lines depict the 2
SD bands.}
\label{gp-profile}
	\vspace{-10pt}
\end{figure*}
\noindent
The GP (for Gaussian process)  profiles are based on a distribution of continuous
log$_2$ intensity profiles for control and treated samples for each
transcript using Gaussian processes (described
in~\cite{vertes-etal-2018cmsb}). Arrays of means and standard
deviations (SD) for control and treated processes were generated by
sampling the continuous distributions at 100 time points between 0
and 48 h chosen uniformly on a log scale.\footnote{The
distribution of intensity profiles projects to a Gaussian
distribution with associated mean and standard deviation at each
time.}
The ratio was then computed as the difference between the treated
and control ($log_2(X/Y) = log_2(X) - log_2(Y)$). Maps
associating transcripts to a number proportional to the space
between the $1$ or $2$ SD bands around the control and treated
means were defined. An approximation of the derivative of ratio time profile was computed as another view of the data.
Figure \ref{gp-profile}  shows GP profiles for
four transcripts with different features.
The control 2 SD bands for
CYP1B1, EGR1 and TXNIP are fairly narrow,  and the control
is basically flat for CYP1B1 and EGR1 while it is rising
for TXNIP and has a small downward slope for CCNB2.  
CYP1B1 has substantial white space between the 2 SD bands 
during the middle times, while EGR1 has
substantial white space between the 2 SD bands
at the later times, indicating a significant change.
TXNIP has no  white space between the 2 SD
bands, but does have white space between the 1 SD
bands at 15 time points (computed from the control and treated SD profiles).  \omitthis{
Interestingly, the CCNB2 control is very
noisy, while the treated is much more `organized' and there
is a bit of white space at latest time.}

\paragraph{PCA and cluster analysis.}
Principle Component Analysis (PCA) (based on
ratio and on derivative GP profiles) was used for dimension reduction
and to rank transcripts according to their contribution to overall variation.  For the methamidophos data, three PCA components account for $95\%$ of the variation. We
explored several standard clustering algorithms. In this
study, we focused on $k$-means clustering with 128 clusters, as this clustering method seems to give the most useful information.\footnote{This is based mainly on size of clusters and distribution of cluster sizes.}
Clusters were computed (for both ratio and derivative) using the PCA 
3-dimensional representation of GP profiles.
Clusters were annotated with GO process, GO function,  
and HUGO Gene Nomenclature Committee (HGNC) family
terms using a heuristic method to compute likely
classifications of a cluster from classifications of its
elements.~\cite{vertes-etal-2018cmsb}


\paragraph{Ranking.}
In addition to PCA ranking, autoencoder techniques were used to derive
compressed representations of the GP profiles. Contrary to the usual
criteria, the transcripts that are least accurately recovered are the
potentially interesting ones, since they don't behave like most
transcripts, i.e. they are anomalies from the autoencoders perspective
(see~\cite{vertes-etal-2018cmsb}).
Generative Adversarial Network (GAN) techniques were used
define two additional rankings: atypical/fake and typical/real.
The GAN ranking algorithm is described in the next subsection.

\subsection{Generative Adversarial Networks}\label{mm-gan}

Generative Adversarial Networks (GANs) \cite{gans} use a
game-theoretic formulation to define the objective function of a
machine learning problem. They involve two subnetworks, a
\textit{generator} and a \textit{discriminator} that play against each
other, so that ideally improvements in one network during training
lead to corresponding improvements in the other until a Nash
equilibrium is reached.  Advantages of GANs are that they learn a
generative model that approximates the full data distribution and they
are less prone to overfitting than pure discriminative approaches due
to the lack of a direct coupling between the generative model and the
data. While there are other interesting generative applications (see
e.g. \cite{canes}), we are primarily interested in obtaining a robust
discriminator to distinguish time series anomalies from typical
behavior.

Our training data set consists of $90\%$ (the other $10\%$ are used
for validation) of the log-ratio time series (excluding the
uninformative time $0$) for all genes estimated using Gaussian
processes as explained in \cite{vertes-etal-2018cmsb}. The data is locally
normalized for each time point (to mean $0$ and unit variance) to
abstract from the absolute magnitude.  Our discriminator network with
an input dimension of $100$ (size of the time series) consists of two
layers, a 1D convolutional network (we use $20$ features, window sizes
$21$, $41$, and $61$, stride $1$) with 
exponential linear unit (ELU) activation, followed by a
dense layer with linear activation. Note that we deviate from the
traditional convolutional architecture by not using a pooling layer
for better stability of the training process. The generative network
takes a $20$-dimensional normal distribution as an input, that is
passed through a dense layer with ELU-activation resulting in a tensor
of dimension $80 \times 20$, $60 \times 20$, or $40 \times 20$,
depending on the window sizes discussed above. Finally, we apply a transposed 1D
convolutional layer to obtain a time series represented as a vector of
dimension $100$. The GANs are specified using TensorFlow~\cite{tensorflow} and Edward~\cite{Tran17}, a framework for
probabilistic programming that recently became part of TensorFlow
Probability.\cite{tfprobability}
We use TensorFlow's Adam optimizer with default settings with the
standard GAN loss function. Each GAN is trained for $100000$ epochs
with a batch size of $1000$ and achieved good convergence (based on
the loss on the validation set) for all the window sizes. A second set
of GANs with the same window sizes was trained on a globally normalized
data set (over all time points and genes, thereby maintaining relative information about the magnitude
of changes), which required a larger feature dimension of $50$ to
achieve satisfactory convergence.

Finally, we applied the discriminators (for all window sizes and the
two variations in terms of normalization) to the full set of log-ratio
time series to obtain a score in the interval $[0,1]$ and a
corresponding ranking of all genes, with values closer to $0$ denoting
an anomaly (fake) and values closer to $1$ denoting a more typical
(real) element of the overall time series distribution. The rankings
based on globally normalized training data turned out to be more
useful in our analysis, in that transcripts that are up/down regulated
by 1 log$_2$ fold or have substantial 1 SD separation are more likely
highly ranked (appear at the extremes of the ranking list) than
transcripts that exhibit much smaller change and/or less separation of
treated from control. Thus we used only the ranks based on globally
normalized data in our analysis process.

\subsection{Siamese Convolutional Networks}\label{mm-twin}

As discussed in more detail in~\cite{canes} we decompose the problem
of causality detection into two subproblems, each addressed by a type
of Siamese neural network.\cite{Bromley94} The first model is
designed to detect and probabilistically quantify the existence of
causality, while a second model is used to probabilistically determine
its direction.

The \textit{undirected causality detector} is a Siamese neural
network, that is a neural network with two identical subnetworks that
are each responsible for processing one of the arguments of the binary
symmetric causality relation. Each argument is a time series (of size
$w = 80$ consistent with our synthetic data to be explained
below). The replicated subnetwork is a 1D convolutional network (we
use a window size $w' = 61$ and stride $1$) with bias and a
rectified linear unit (ReLU) 
activation function yielding a tensor ($20 \times 50$
dimensional). This is followed by an average pooling layer yielding a
vector ($50$ dimensional), followed by a dense layer with the same
output dimension and again with bias and relu-activation. Hence, the
output of each subnetwork is a vector in feature space ($50$
dimensional). The two outputs are combined by a dot-product layer
(which captures the symmetry of the problem as part of the
architecture). As usual for Siamese networks, the symmetry is also
exploited by weight-sharing between the two subnetworks. Finally, to
obtain a probabilistic output, a sigmoid function is applied to a
linear transformation of the scalar result from the dot-product, which
can also be viewed as trivial dense layer with bias and
sigmoid-activation. As a loss function we use binary cross-entropy and
as an optimizer we use TensorFlow's implementation of Adam (with
default parameters).

The training and validation data set is based on a synthetic dynamic
model, that we can only briefly summarize here and refer to
\cite{canes} for more details. Using our Gaussian process as a
template we generate positive and negative examples of causally
related ``synthetic'' genes, or more precisely, corresponding time
series that are biological plausible in the context of our experiment.
There is a complexity parameter $m$ involved in the construction that
we refer to as mixin-parameter. It allows us to vary the complexity of
the synthetic model by limiting the number of genes that can
participate in an interaction, which is modeled by a noisy linear
superposition of time series sampled from our Gaussian processes model
with a variable time lag. Employing curriculum training, we train the
models with increasing complexity of the training data set. We use the
mixin-parameter with settings $m = 0, 2, 4$ to vary the complexity of
the synthetic model. While we also explored higher parameters, we
estimated that a target of $m = 4$ should be sufficient for
biologically plausible results.
Each stage is
trained for $1000$ epochs with a batch size of $20000$. Our synthetic
data set contains $2000000$ pairs, from which we generate training and
validation sets by random split of $90\%$ vs. $10\%$.

To detect the direction of an already established causality, we define
a \textit{lag detector} again a Siamese neural network, albeit a
somewhat unconventional one detailed in \cite{canes}, 
that after training estimates the lag in a normalized range $[-1,1]$. 

As before, the replicated subnetwork is a
1D convolutional network with bias and a ReLU-activation function
followed by a dense layer again with bias and ReLU-activation. Unlike
in the causality detector there is no pooling layer involved. Hence,
the output of each subnetwork is a tensor ($20 \times 50$
dimensional). After flattening, the two outputs are combined by a
subtraction layer (which captures the antisymmetry of the problem)
yielding a vector (of dimension $1000$), The next layer is a dense
layer without bias and tanh-activation reducing the dimension to $50$,
and finally a linear dense layer without bias is used to obtain a
scalar output. As a loss function we use mean square error and as an
optimizer we again use TensorFlow's implementation of Adam (with
default parameters). The training and validation data set for the lag
detector is generated in the same way as the (positive) synthetic
pair set for the causality detector (again $1000000$ pairs), but we
allow and track positive, negative, and zero time lags in the
construction. Training and validation sets are then generated from
this set of labeled pairs.  For training we again use curriculum
learning with the same parameters as for the causality detector.

We use these Siamese networks to synthesize two types of
graphs. An \textit{undirected causal network} is defined by nodes
corresponding to all genes in a given subset and edges between pairs of
genes for which the causality detector detects a dependency with at
least the cutoff probability \longversion{(another parameter in the graph
synthesis process)}. A \textit{directed causal network} is a refinement
of an undirected causal network. Each edge is directed according to
the prediction of the lag detector based on a positive threshold in
the interval $(0,1]$ that was associated with a probability during
validation. Each undirected edge becomes a directed edge if the lag
prediction reaches at least the positive threshold and remains
undirected otherwise.

As an example, the validated accuracy for the
synthesized networks used in this paper 
using a synthetic model with mixin-parameter $m = 4$) is
$0.75$ for the existence of a causal relation and $0.72$ for its
direction.

\subsection{Time Warp Causal Inference}\label{mm-bayes}
\emph{Time warp causal inference} employs two primary algorithms for its operation:  bootstrap resampling of the data and alignment of cause and effect events in time using the Needleman–Wunsch algorithm.\cite{needleman-wunsch-1970jmb} 
Comparison with N-th order, dynamic Bayesian networks indicates superior performance for time warp causal inference.

We begin by setting a threshold for significant fold change – a threshold of $2.0$ is common.\footnote{Note that here we are using linear fold change rather than log$_2$.}  A fold change over $2.0$ is considered high; below $1/2$ is considered low.  For every transcript at every time point, there are two sets of concentration data:  a treated set and a control set.  The statistical t-test determines whether the means of the two sets are statistically different with a given confidence level.  If the t-test indicates that they are different with satisfactory confidence, we can compute mean fold change (treated/control) to determine if the transcript is high or low at that point.  Thus, the t-test can distinguish among three cases:  high, low, and not statistically different.  We need to distinguish among four cases:  high, low, same (nominal), and too noisy to make a conclusion with high confidence.  Data points that are too noisy may be treated as missing.

\omitthis{
\begin{figure}[h]
\centering
\includegraphics[width=0.5\textwidth]{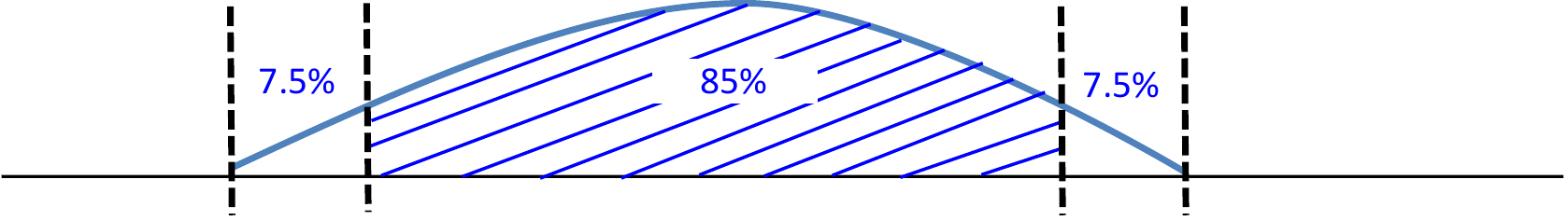}
\caption{Distribution of 500 mean fold changes from bootstrap resampled data.  To provide $85\%$ confidence level, the highest and lowest $7.5\%$ of the distribution are ignored.}
\label{andy-fig1}
\end{figure}
}
We are able to make the distinction among the four cases by using bootstrap resampling of the data.  For a specific transcript at a given time point, assume each of the two data sets contains 3 concentration readings.  With these two data sets, we can calculate 9 possible fold change ratios and then to calculate a mean fold change ratio for the data set. To bootstrap resample:
\begin{itemize}
  \item	Repeat 500 times
  \begin{itemize}
   \item
		Sample with replacement the three readings in the treated set to make a new
treated set
	\item
  Sample with replacement the three readings in the control set to make a new
control set
  \item
  calculate the fold change ratio mean using the two, new data sets
  \end{itemize}
  \item
  	Sort the $500$ means to make a fold change distribution for this
     transcript at this time point
\item Based on the desired confidence level (e.g. $85\%$), 
   delete the extremes of the distribution
 \item
  Compare remaining distribution with fold change thresholds
\end{itemize}

\begin{figure}[ht]
\centering
\includegraphics[width=0.7\textwidth]{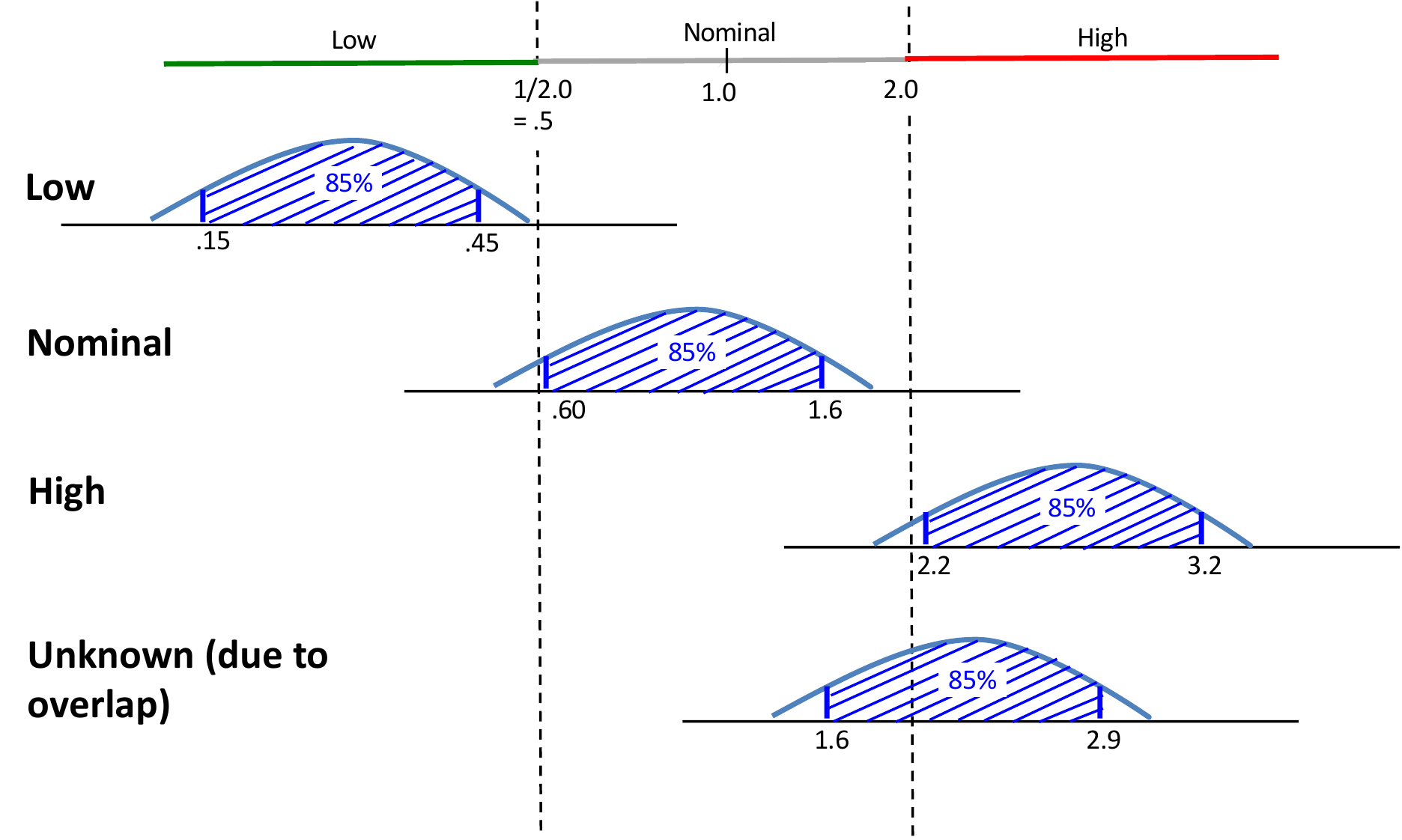}
\caption{In the first distribution, the central $85\%$ of the distribution all lies below the low threshold of $0.5$, so the distribution is labeled low.  In the second distribution, the central $85\%$ is between the low threshold and the high threshold of $2.0$, so this distribution is labeled nominal (treated and control are same).  In the third distribution, the central $85\%$ all lies above the high threshold, so the distribution is labeled high.  In the last distribution, the central $85\%$ overlaps both the nominal area and the high area, so this distribution is labeled unknown.}
\label{andy-fig2}
\end{figure}

\begin{figure}[h]
	\vspace{-15pt}
	\centering
	\includegraphics[width=0.5\textwidth]{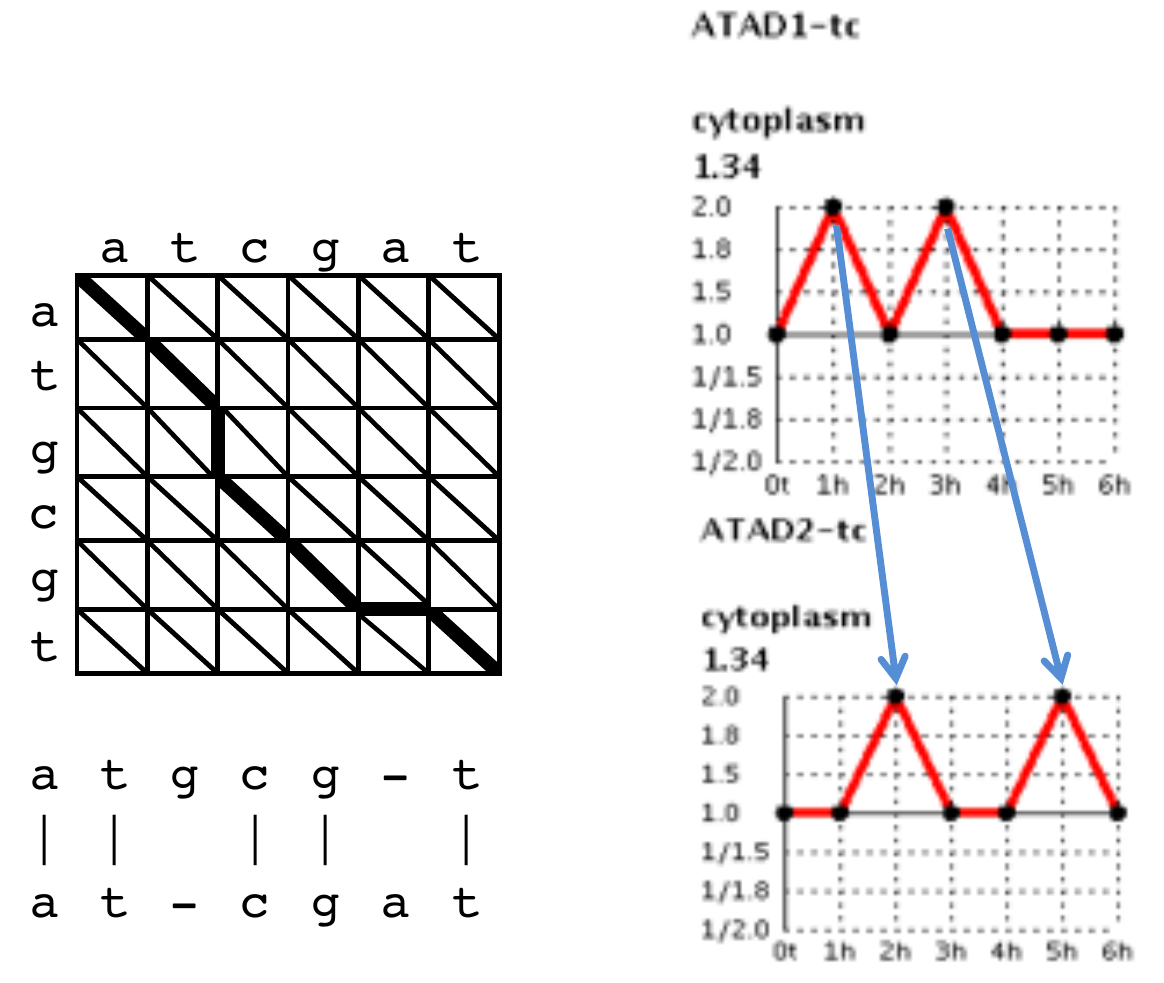}
	\caption{Timewarp illustration.}
	\label{andy-figs-3-4}
	\vspace{-15pt}
\end{figure}

\noindent
The Needleman-Wunsch alignment algorithm~\cite{needleman-wunsch-1970jmb} is widely used for global, genetic sequence alignment.  This algorithm uses an edit graph to align two genetic sequences, performing approximate matches to subsequences in arbitrarily different locations in each sequence by warping (adjusting) the space between the two sequences using insertions and deletions as appropriate. The algorithm uses dynamic programming that minimizes a cost function reflecting the cost of the insertions and deletions.  
To align cause and effect events with unknown, variable delays between cause and effect, time warp is needed (analogous to the way Needleman-Wunsch warps space in sequence alignment).  To make our time warp edit graphs, each row and column of the edit graph is labeled with a value (high, nominal, or low) and the time of measurement.  Unknown values are omitted.  We then align the two event sequences using dynamic programming, preserving causality:  effects events cannot precede cause events.  A slightly higher cost is associated with aligning two events that are widely separated in time.  Since causes can activate or inhibit effects, we initially match non-nominal events without regard to direction (high or low).  Once the optimal alignment is identified, we produce both activate and inhibit alignment scores.  For analysis purposes, we use the larger of the two scores exclusively.
Figure~\ref{andy-figs-3-4} illustrates the algorithm. On the left,
the subsequence “cg” is matched by the Needleman-Wunsch algorithm despite being offset in the two global sequences. 
\omitthis{The “-“s represent insertions or deletions in the global sequence required for the optimal sequence alignment.}
On the right, the time warp causal inference algorithm matches both event pairs even though they have differing delays.  N-order dynamic Bayesian networks are unable to find such matches.

\section{Results of data analysis}\label{results}

\subsection{Global response features}\label{global}

\paragraph{Temporal distribution of response.}
To get an idea of the shape of SK-N-AS cell response to methamidophos treatment we compute maps from transcripts to the first measured time when the log$_2$ ratio at that time is greater than $1$ (up by map) or less than $-1$ (down by map).

\begin{figure}[h]
\vspace{-15pt}
\centering
\includegraphics[width=0.9\textwidth]{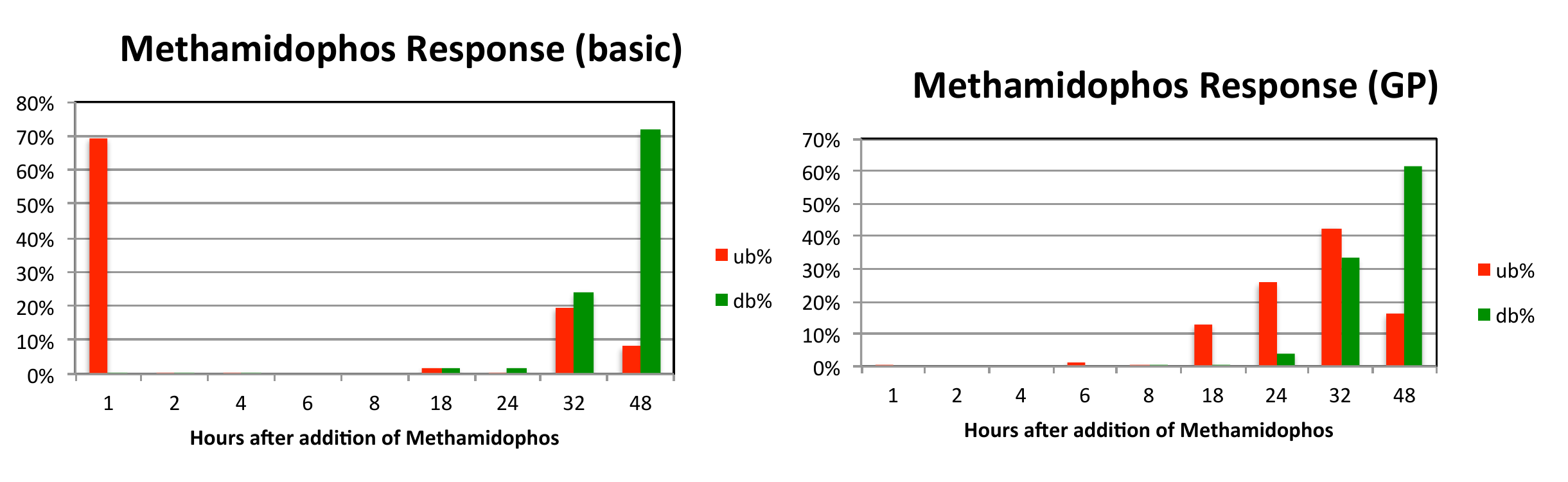}
\caption{ Up/Down By Charts showing
the percent of transcripts up/down by (first crossing the 1 log$_2$ fold threshold) at each measured time point. (Left) computed using the 
basic time profiles. (Right) computed using the GP time profiles. }
\label{udby}
\vspace{-15pt}
\end{figure}

\noindent 
The results are summarized in Figure~\ref{udby}. The basic chart
(left) shows a large spike of upregulation at 1h. Many of these
transcripts show a peak at 1 h and little change at other times.
Volcano plots (not shown) for the transcriptomics data suggest a
possible experimental artifact at 1 h. This suggests the 1 h
responders should be viewed with suspicion. This is smoothed away in
the GP chart (right). Conversely, the basic chart shows little further
action until 32 h while the GP chart shows substantial upregulation
beginning at 18~h. This suggests that it is important to consider
analysis based on both the basic time profiles and the GP time
profiles, and to have multiple criteria for selection, to get the most
out of the data.

\paragraph{Cell cycle.}

\begin{figure}[h]
\centering
\includegraphics[width=0.8\textwidth]{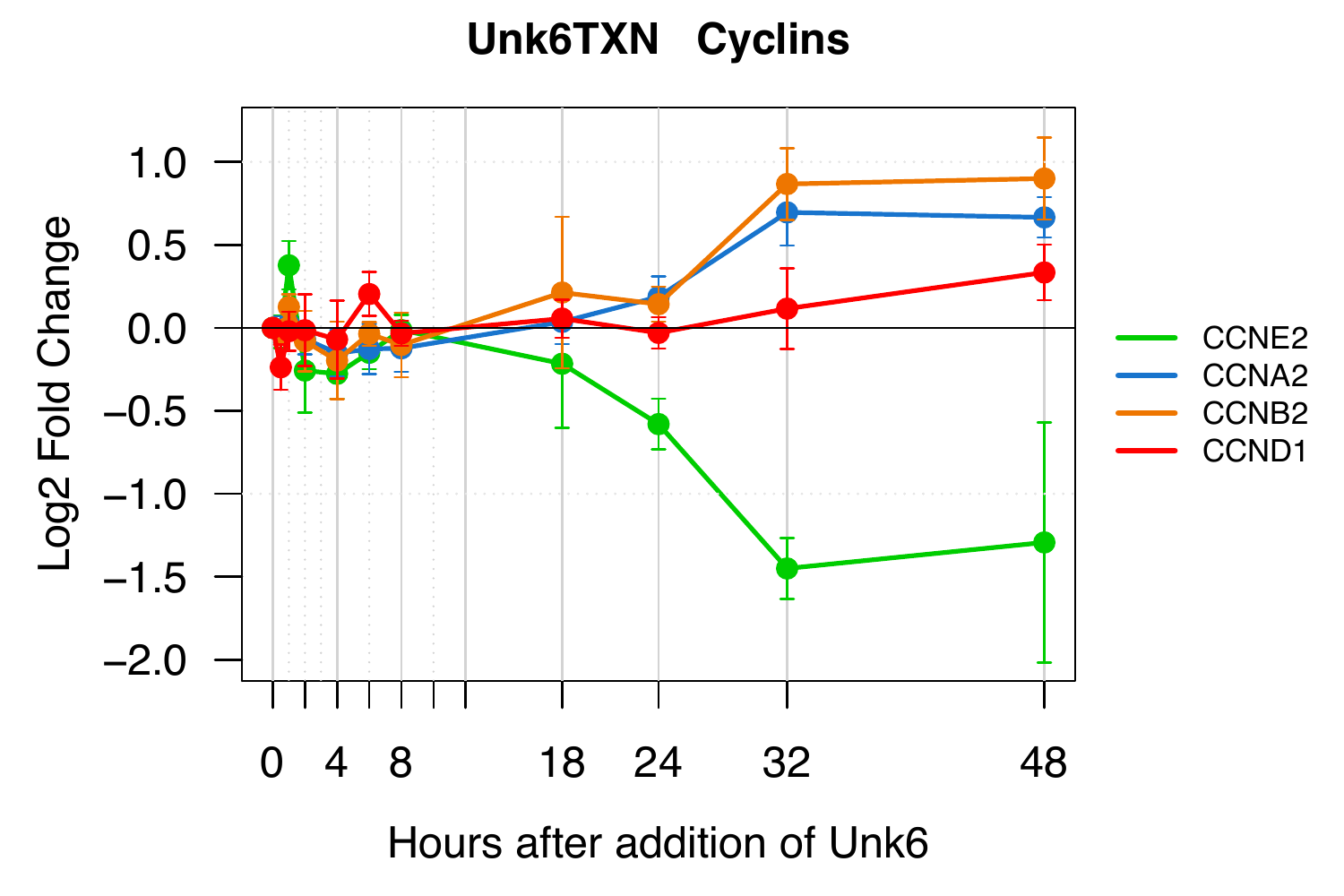}
\caption{Cell-cycle state: Basic time profiles for Cyclins}
\label{cyclins}
\end{figure}

We used the cyclin cell-cycle checkpoint markers CCNE2, CCNA2, CCNB2,
and CCND1 to get a picture of the global cell cycle state
of the cells after methamidophos exposure. 
Figure~\ref{cyclins} shows the basic time profiles of these transcripts.
We see a down regulation of CCNE2 from 18 h, indicating that
the cells have has passed G1/S.  There is a 
sustained upregulation of CCNA2 and CCNB2 starting at 32 h,
indicating a G2 arrest at 32 h. 

\vspace{-5pt}
\paragraph{Predicates.}

We used the various ranking functions and significance filters to define
predicates to select sets of transcripts to examine.
Predicates \emph{upby t} / \emph{dnby t} 
are derived from the up/down chart maps.
For $t$ one of the measured times, a transcript satisfies
\emph{upby t} (\emph{dnby t}) if the log$_2$ ratio first passes $1$ ($-1$) at
at time t. 
Thus for $t = 18 h$, \emph{upby t} is the set of
transcripts with log$_2$ ratio first exceeding 1 at 18 h. 
We defined a
modest size (194) set of transcripts, named \emph{Top20X}, 
to initially consider.
Top20X consists of transcripts that are ranked in the top 20 of at
least one of the ranking functions: PCA (ratio or derivative), any of 3
principal components positive or negative; GAN fake (atypical, anomalous) or real (typical), according to one of the three GAN models; or one of the autoencoders-based ranks. Furthermore
transcripts in Top20X must have a (significant) log$_2$ fold change 
magnitude of at
least 1 at some measured time, or have 1 SD separation of
control and treated GP profiles for at least 15 (of 100) sampled
times.\footnote{Causal network information was not used in the definition of Top20X. More details about these predicates can be found
in~\ref{predicates}. A spread sheet listing these transcripts and many
of the properties of their time profiles can be found at
\cite{sortable}.}

We note that of the transcripts upregulated by 1 h (recall the 1 hour spike in Figure~\ref{udby}) only 24 belong to Top20X. Of these, only
TIPARP has sustained upregulation.  It is classified GAN real.
The rest are included in Top20X because of PCA ranking. This seems to
confirm that although the 1 h data is suspect, perhaps it should not simply be dropped.

\subsection{Candidate MoA Biological Processes}\label{processes}

To identify cellular processes that are part of the
SK-N-AS cell response to methamidophos, we examined
GO process terms associated with responsive transcripts
in several ways.
PantherDB\cite{mi-etal-2017panther} was queried for over representation using several  transcript sets
(see Appendix \ref{panther} for details).
For example, from the 545 transcripts upby 1 h, interferons
were identified in an immune response group.  
From 155 transcripts upby 32 h,
multiple process groups were found, including:
metabolic processes, response to growth factors, circadian rhythm,
chromatin related, and cell cycle.  From 62 transcripts
upby 48 h, heat shock and protein folding stood out.
We also manually scanned GO process terms associated with clusters by
our cluster analysis algorithm, and GO process terms associated with
the Top20X transcripts by UniProt.
Three process types stood out: unfolded
protein response (UPR, including ER stress), cyclic adenosine
monophosphate (cAMP) response, and calcium ion (Ca++) related processes.
In addition, analysis of the data with the MIRU
tool\cite{wright-etal-2014biolayout3d} found a cluster associated
to cell-cell signaling, suggesting we look for 
cell-cell signaling indicators in our analysis (see Appendix \ref{cell-cell}
for details).
Using the resulting process suggestions we investigated further.

\paragraph{Unfolded Protein Response (UPR).}

Five transcripts were identified by Panther overexpression analysis
of the upby 48 h transcripts as being annotated with UPR: HSPA1A,
HSPA1B, HSPA5, HSPA6 and DNAJB1. The first four code for heat shock
proteins. DNAJB1 stimulates the folding of
unfolded proteins mediated by HSPA1A/B~\cite{PubMed:24318877}.\hfill\break

\begin{figure}[h]
	\centering
	\includegraphics[width=0.8\textwidth]{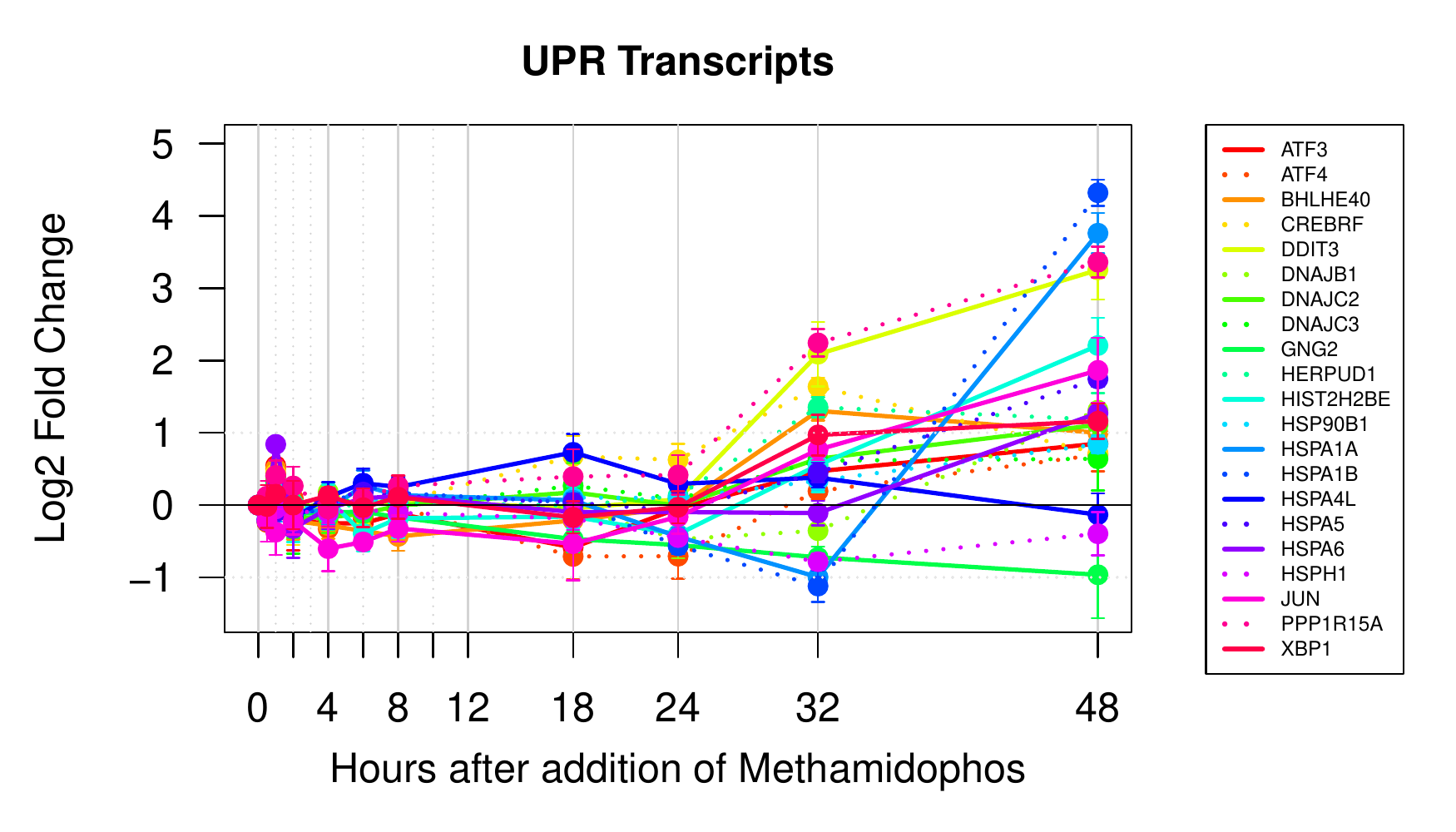}
	\caption{Time profiles of UPR  transcripts.}
	\label{upr}
\end{figure}


From six  clusters annotated with UPR related terms, nine 
transcripts were annotated with UPR related terms: 
BHLHE40, DNAJC2, HIST2H2BE, HSP90B1, HSPA4L, HSPA5, HSPH1, JUN,  and XBP1. 
Five additional Top20X transcripts are annotated with UPR terms:
CREBRF, DDIT3, GNG2, HSP90B1, HSPA4L, HSPA5, JUN, and PPP1R15A. 
Taken together we have seventeen
transcripts associated to UPR based on our data
analysis.  Five are ranked in the top 20 GAN real (DDIT3, HIST2H2BE, HSPA5, JUN, PPP1R15A) and two are ranked in the top 20 GAN fake (CREBRF and GNG2).
We also consulted the Reactome ``Genes involved in Unfolded
Protein Response'' pathway and obtained 3 additional transcripts to
consider: ATF3 DNAJC3 and HERPUD1. Finally, the paper
\cite{PMID:23563539} shows that the gene for ATF4 is upregulated  
at 12-24 h in response to three out of seven
drugs commonly used as UPR inducers.
In all, twenty one transcripts from Top20X were found to be related
to UPR using GO annotations, pathway databases, and experimental results. 
Figure~\ref{upr} shows the overlaid basic time profiles for 
these twenty one transcripts.
For most of the transcripts, upregulation seems to start between
24-32 h, while HSPA1A and HSPA1B are downregulated 18-32 h
and upregulated 32-48 h.
Both GNG2 and HSPH1 are downregulated from 32-48 h.

\paragraph{Cyclic Adenosine Monophosphate (cAMP) response.}

The cAMP-dependent pathway is a G protein-coupled
receptor-triggered signaling cascade, activated by
a number of toxins, with a role in many 
biological processes, including cell
communication~\cite{wiki1}. cAMP is produced by conversion of ATP
by activated adenylate cyclase (also called adenylyl cyclase).
The diterpene forskolin is used in experiments to study cAMP response, 
as it
directly activates adenylate cyclase thus increasing the level of cAMP. 
Using time series transcriptomics data from 
HepG2/C3A  cells treated with forskolin~\cite{vertes-etal-16hupo}
we collected lists of transcripts up or downregulated early as a model of cAMP responsive genes.

\begin{figure}[h]
\vspace{-5pt}
\centering
\includegraphics[width=1.0\textwidth]{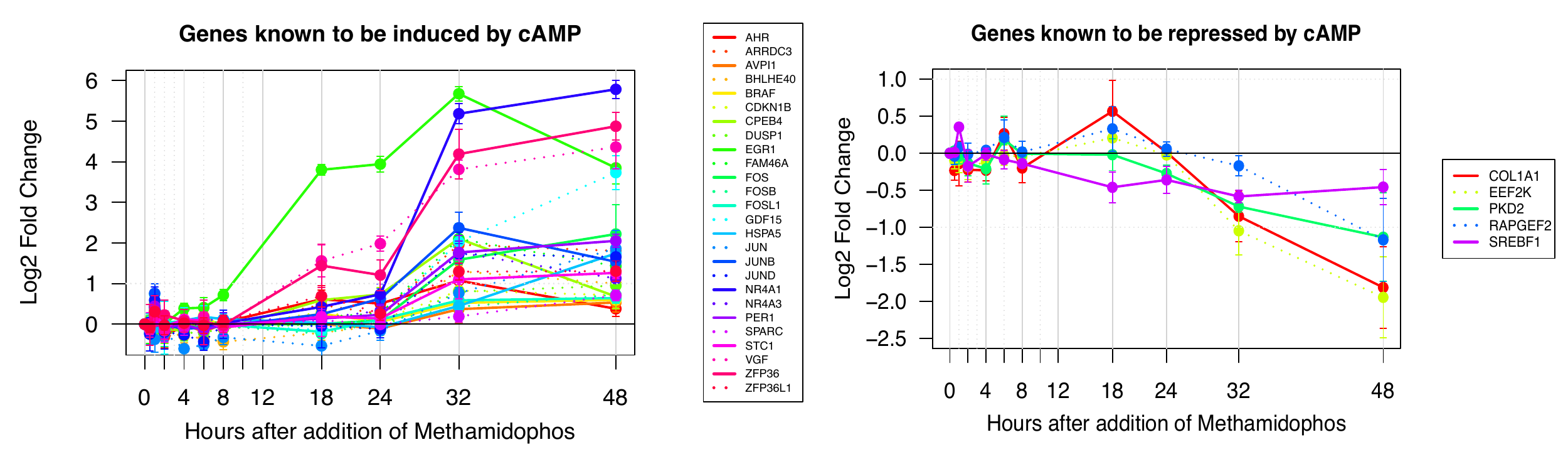}
\caption{Time profiles of cAMP transcripts}
\label{cAMP}
\vspace{-5pt}
\end{figure}
\noindent
Figure~\ref{cAMP} shows plots of the time profiles of forskolin responsive transcripts that also show some
response to methamidophos.
\longversion{
Except for EGR1 (solid green) and the ZFPs (reddish) the up
regulation appears to start around 24 h. The down regulation seems
to begin around 18 h. This is consistent with a first phase of up
regulation that produces something that activates adenylate cyclase.
}
Of the twenty six upregulated transcripts suggested by the
forskolin data, twelve are in the methamidophos Top20X: ARRDC3, BHLHE40, CPEB4,
EGR1, FOS, GDF15, HSPA5, JUN, NR4A1, PER1, VGF, and ZFP36. 
\longversion{Only five
of these transcripts (FOS, HSPA5, JUN, PER1, and VGF) were in our
set of cAMP annotated transcripts. On the other hand, ID4 and RELN
were cAMP annotated and are not in the forskolin list.}
Of the five 
downregulated transcripts suggested by the forskolin data, two are in  the methamidophos Top20X: COL1A1 and EEF2K. 
\longversion{Both are cAMP annotated.}

\vspace{-5pt}
\paragraph{Calcium ion response.}\label{calcium-ion}
Using UniProt GO process annotations we found twenty one Top20X transcripts annotated with calcium ion (Ca++) related terms.

\begin{figure}[h]
\centering
\includegraphics[width=1.0\textwidth]{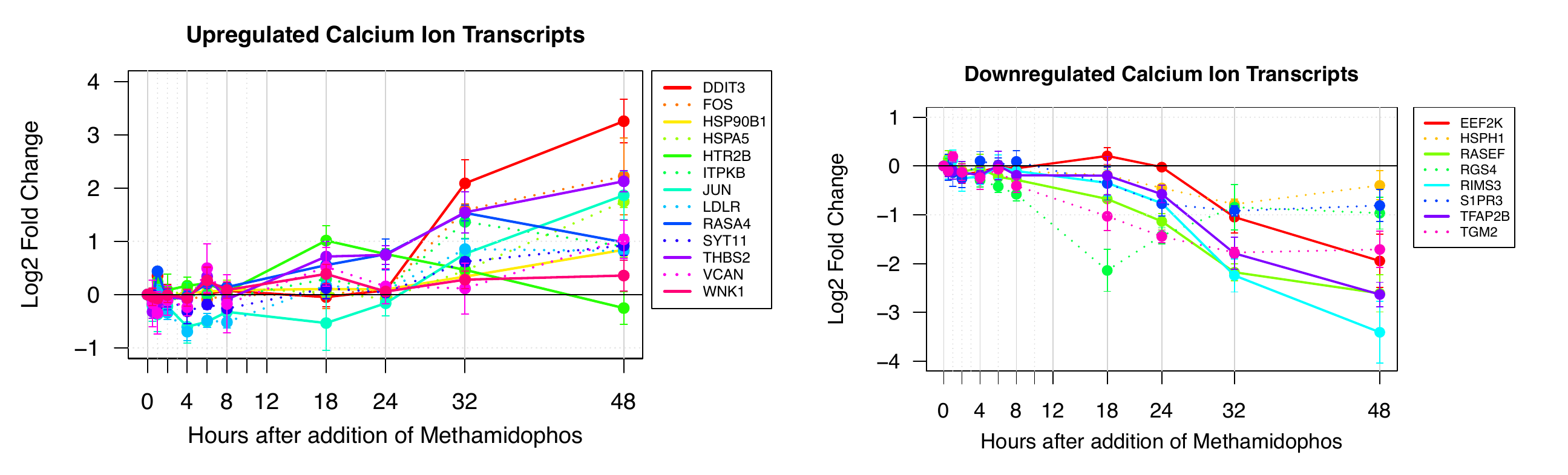}
\caption{Profiles of calcium ion related transcripts.}
\label{calcium-ion}
\end{figure}

\noindent
Thirteen are upregulated (HTR2B up by 18h; DDIT3, FOS, ITPKB, LDLR, RASA4, THBS2, WNK1 down by 32 h; and HSP90B1, HSPA5, JUN, SYT11, VCAN up by 48 h) and eight are down regulated
(RGS4, TGM2 down by 18 h; RASEF, RIMS3, S1PR3 down by 24 h; EEF2K, HSPH1, TFAP2B down by 32 h).
Six of the upregulated and four of the down regulated Ca++ transcripts are ranked GAN real.
Two of the upregulated Ca++ transcripts are ranked GAN fake.
Figure~\ref{calcium-ion} shows the expression profiles of these
transcripts: upregulated on the left and downregulated on the right.

\omitthis{
(DDIT3, FOS, HSP90B1, HSPA5, HTR2B, ITPKB, JUN, LDLR, RASA4, SYT11, THBS2, VCAN, WNK1)

up 13
HTR2B 18
DDIT3 32  GR  upr      dd cc
FOS 32    GR      cAMP 
ITPKB 32  GF           
LDLR .8@32 -
RASA4 32  GF           
THBS2 32  TM 
WNK1  32   -
HSP90B1 .8@48 upr       hypox
HSPA5   48  GR  upr  cAMP
JUN     48  GR  upr  cAMP cc  h2o2
SYT11   .9@48 -
VCAN    48     -

(EEF2K, HSPH1, RASEF, RGS4, RIMS3, S1PR3, TFAP2B, TGM2)

dn 8
RGS4  -18  GR
TGM2  -18  GR-
RASEF -24  GR
RIMS3 -24  GR-
S1PR3 -.9@24 -
EEF2K -32    -    cAMP
HSPH1 -.7@32 -
TFAP2B -32 -
}

\subsection{Conjectured acetylcholine build up}

\begin{figure}[h]
\centering
\includegraphics[width=0.8\textwidth]{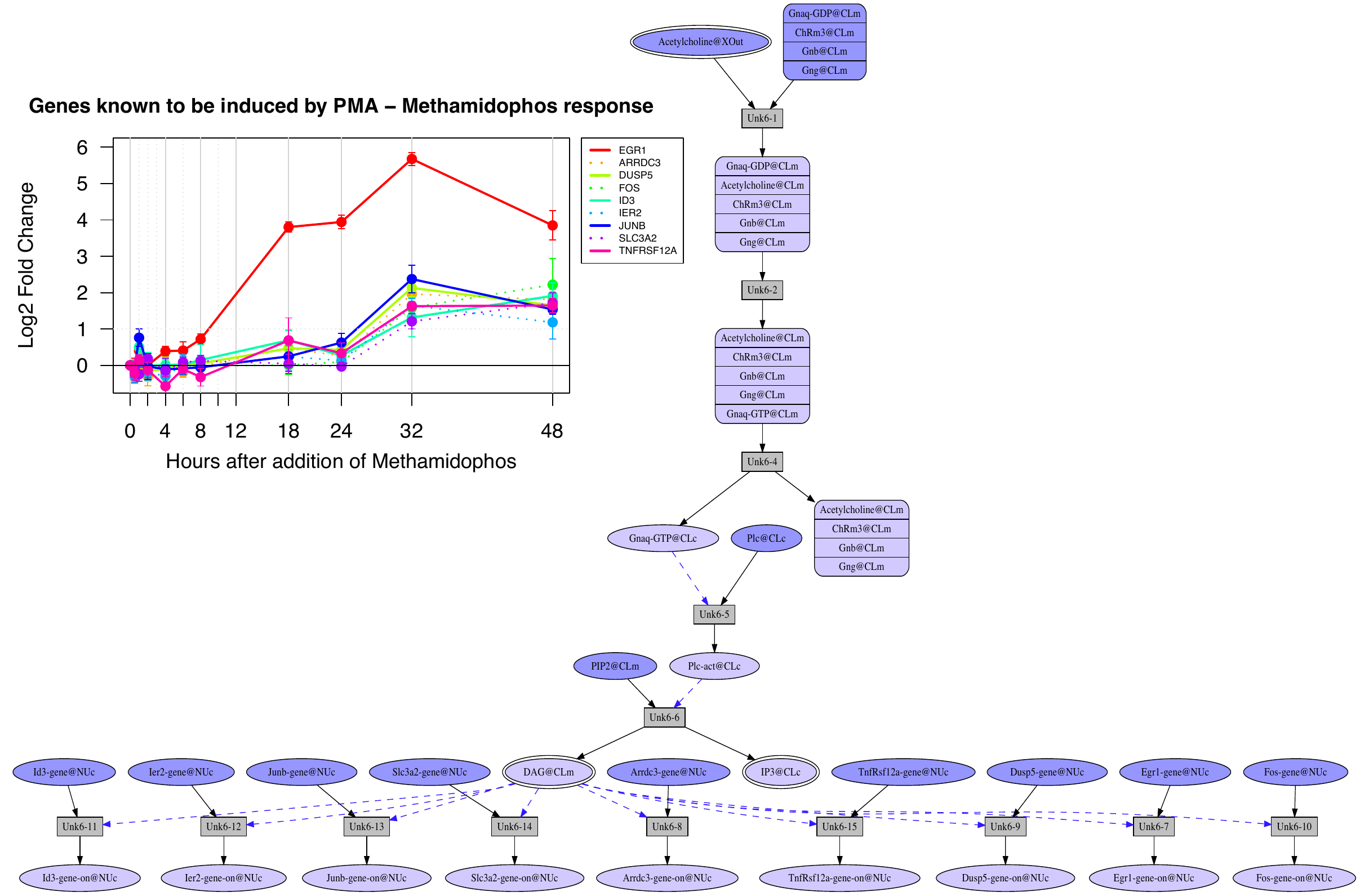}
\caption{Response to acetylcholine buildup}
\label{pma-dag}
\end{figure}

\noindent
The result of acetylcholine esterase inhibition in cultured cells is that acetylcholine builds up and binds and activates muscarinic and nicotinic acetylcholine receptors~\cite{pmid24179466}.
Since methamidophos is a cholinesterase inhibitor we looked for the
consequences of a buildup of acetylcholine.
Figure~\ref{pma-dag} shows a curated pathway leading from the binding
of acetylcholine to the muscarinic receptor ChRm3 to 
upregulation of genes known to be upregulated by PMA, a 
diacylglycerol (DAG) mimic. 
This pathway corresponds to the production of DAG and IP3 via 
break down of PIP2 by activated PLC.
We used the Pathway Logic Datum knowledge base\cite{datumkb},
the derived PMA response network~\cite{STM8online}
and induction information from UniProt 
to identify the genes responding to PMA.
The inset in Figure~\ref{pma-dag} shows the methamidophos response time profiles of these transcripts.
Note that except for EGR1, these transcripts all are upregulated
starting sometime after 24~h. 
EGR1 responds to many signals,
what is possibly surprising is its delayed upregulation, as it
is one of the immediate early genes.
In addition to the DAG induced response, IP3 
initiates intracellular calcium release, consistent with
the Ca++ response noted above.

\begin{figure}[h]
\centering
\includegraphics[width=0.8\textwidth]{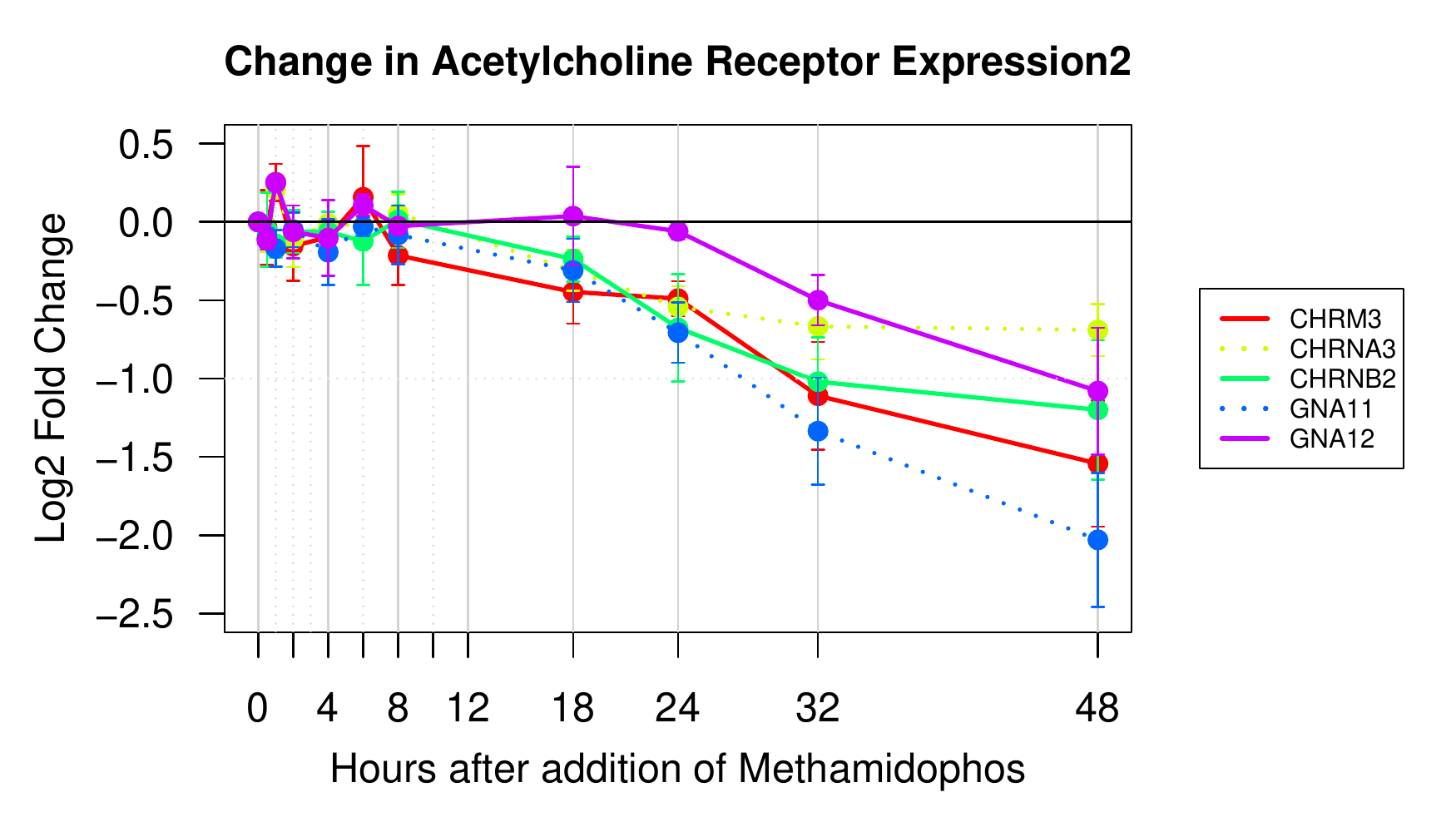}
\caption{Down regulation of Acetylcholine receptors}
\label{achr-dn}
\end{figure}
\noindent
The acetylcholine receptors 
CHRM3, CHRNA3, and CHRNB2, 
and GPCR binding partners components GNA11, GNA12 are downregulated
starting around 18 h (see Figure \ref{achr-dn} ).   This could be additional evidence of
acetylcholine build up, with the cell damping the response.

\vspace{-5pt}

\paragraph{Overlapping roles.}
There is non-trivial overlap between
some pairs of process specific transcript lists.  In
particular, ARRDC3, EGR1, FOS, and JUNB belong to
the PMA and cAMP  lists;
BHLHE40, HSPA5, and JUN belong to the UPR and cAMP  lists; 
DDIT3, HSP90B1, HSPA5, HSPH1, and JUN belong to the UPR and 
Ca++ lists; and  
FOS, HSPA5, and JUN belong to the cAMP and Ca++ lists.
The UPR and PMA lists do not overlap.  HSPA5 and JUN belong
to all lists except PMA and FOS belongs to all lists except UPR.
Even though there are overlaps, each of the four process lists has
many transcripts unique to that list.

\subsection{Some distinguished transcripts}\label{distinguish}

In addition to identifying processes potentially
involved in SK-N-AS cell response to methamidophos, 
we are interested in identifying
specific transcripts that might play significant roles.  
For this purpose we identified three different Top20X subsets:
two based on GAN ranking combined with $k-$means clustering
and one based on playing a role in multiple processes.
In the following, HGNC transcript names are often followed
by process annotations in parentheses.

\paragraph{GAN fake transcripts.}
In six $k-$means clusters (three upregulated and three downregulated)
at least half of the Top20X transcripts are GAN fake.  
This includes thirty-five of the forty seven Top20X GAN fake transcripts.
The three upregulated clusters (27, 78, 127) are up by 32h.
Cluster 27 contains nine GAN fake transcripts including 
CREBRF (UPR), DIAPH2 (actin filament organization),
HDAC5 (chromatin organization), and RASA4 (Ca++).
Cluster 78 contains five GAN fake transcripts, of which two,
AHNAK and ITPKB, are Ca++ related. 
Cluster 127 contains four GAN fake transcripts including
GJA1 (cell communication by electrical coupling)
and IER2 (cell motility).
Cluster 81, downregulated at 8 h, contains seven GAN fake transcripts
including three associated to cholesterol biosynthetic process: DHCR7, HMGCS1, and NSDHL.
Clusters 20 and 38 are downregulated at 48 h
Cluster 20 contains four GAN fake transcripts including ACAT2 (cholesterol biosynthetic process) and TOE1 (Target of EGR1).
Cluster 38 contains six GAN fake transcripts including 
GNG2 (protein folding, calcium modulating), ARHGAP28 (negative regulation of stress fiber assembly), and NEBL (actin filament binding).

\paragraph{GAN real transcripts.}
The transcripts of eight $k$-means clusters (five upregulated, three
downregulated) were all (with one exception) Top20 GAN real
transcripts. 
Four of these clusters are singletons:
CYP1B1, TIPARP, EGR1, and NEUROG2.
TIPARP, a negative regulator of AHR, is the one transcript up at 1 h with
sustained upregulation. CYP1B1 (toxin metabolic process) is up by 2 h and
stays up. It is likely oxidizing the toxin. 
EGR1 (Early growth response protein 1, PMA, cAMP)  a transcription factor, generally showing an
early response to perturbations, is not up until 18 h. NEUROG2 (E-box
binding transcription factor) is downregulated at 18h. The GAN real
transcripts NR4A1, VGF, and ZFP36 form a triplet cluster upregulated at
18-32 h. This cluster is a subset of the cAMP responsive transcripts. The
transcripts DDIT3 (UPR, Ca++), GDF15 (cAMP), PPP1R15A (UPR), and TFPI2
(extracellular matrix) form a cluster of size four upregulated at 32 h.
TXNIP and RHOB form a doublet cluster downregulated early (2-4 h).
One role of RHOB is actin filament bundle assembly. According to
UniProt, 
TXNIP is known to be downregulated in response to
oxidative stress. 
Txnip, the protein coded by TXNIP inhibits
thioredoxin activity and the proteasomal degradation of Ddit4.
TXNIP also stands out because it is in the top 20 real transcripts
for one of the GAN models and in the top 20 fake transcripts in
another of the GAN models. This dual ranking is rare.
RGS4 (GAP, Ca++), HAND1 (transcription factor)
and POPDC2 (cAMP-binding, regulation of membrane potential) form a triplet cluster downregulated at 18 h. 
POPDC2 is the exception, it is top 30 GAN real ranked. 
Finally, CACNG4 (membrane depolarization) and NCOA5 (glucose homeostasis and negative regulation of insulin receptor signaling)
form a cluster of 2 downregulated at 18 h.
The remaining GAN real transcripts appear in clusters where most
elements are not GAN ranked.
For example cluster
66 has 5 elements, all in the UPR process list, only two (HIST2H2BE and JUN) are top 20 GAN real.

\omitthis{
QUESTION: is membrane depolarization related to Ca++ release?
CACNG4 -- is also pos reg of AMPA receptors -- but this is about Ca++
influx  -- downreg -- slowing influx???
}

\paragraph{Multirole transcripts.}

Thirteen transcripts were selected from the Top20X
because they were annotated by multiple specific GO process terms that stood out in a scan of the UniProt GO annotations or in multiple 
Panther overrepresentation groups.

\begin{figure}[h]
\centering
\includegraphics[width=0.9\textwidth]{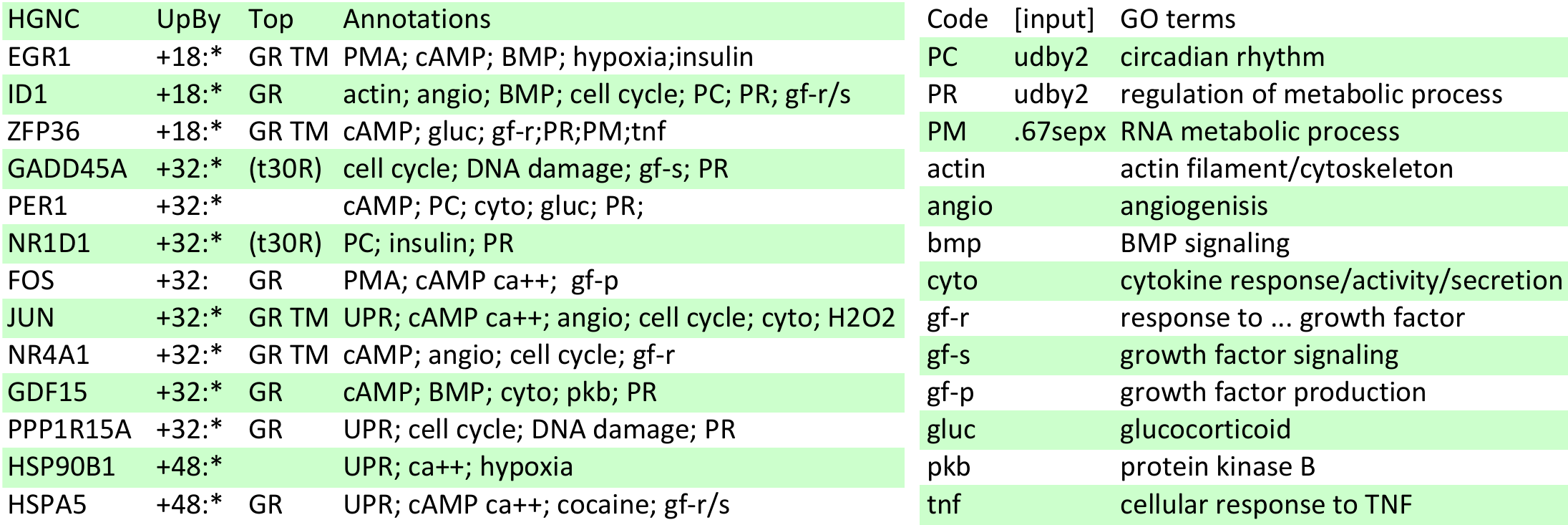}
\caption{Multirole transcripts. 
PMA, UPR, cAMP, and Ca++ refer to the process groups
discussed in section \ref{processes}.
UPR includes unfolded protein response; ER stress, protein folding
Ca++ includes calcium ion release, sequestering and transport.
The P in PC,PR,PM indicate annotations derived from Panther analysis.}
\label{multirole}
\end{figure}
\noindent
The table in Figure \ref{multirole}  summarizes these transcripts 
with their annotations.  
Except for PER1 and HSP901B1 all are top 30 GAN real ranked
and of the latter all but GADD45A and NR1D1 are top 20 GAN real ranked.
Three of the four processes already identified
are well represented: UPR 4, Ca++ 6, and cAMP 8.
Five of the transcripts are annotated with cell cycle.
Three are annotated with BMP signaling, three with
cytokine related processes,  three with angiogenesis
and three with circadian rhythm (PC).
Six are annotated with growth factor related terms and
seven are annotated with regulation of metabolic processes.

\subsection{Inferred causal relations}

\begin{figure}[h]
\vspace{-10pt}
\centering
\includegraphics[width=0.7\textwidth]{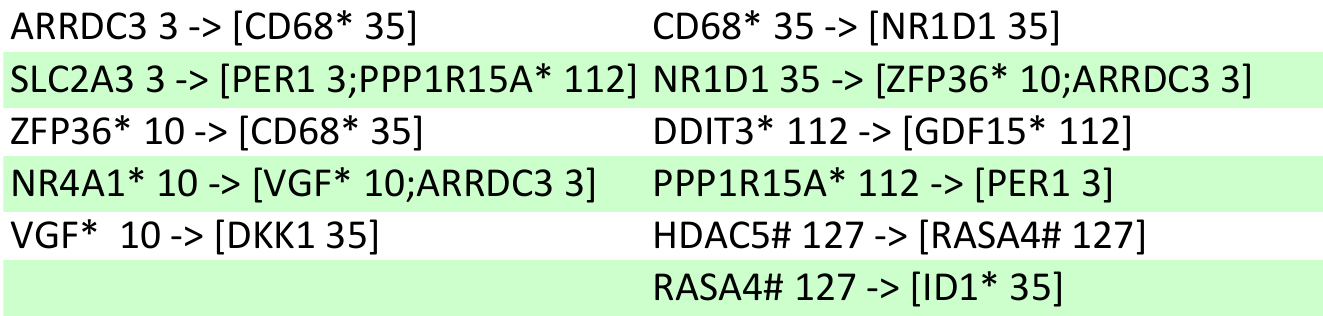}
\caption{Sample Edges from the Siamese Network}
\label{twinedges}
\vspace{-10pt}
\end{figure}
\noindent
Figure \ref{twinedges} shows a sample of the top 200 causal relations proposed by the Siamese Convolutional algorithm  restricted to transcripts in Top20X.  
\texttt{*} indicates top 20 GAN real ranking and
\verb|#| indicates top 20 GAN fake ranking.
The numbers are cluster identifiers.
We see that edges are often within a cluster,
and clusters 3, 10, 35, 112, 127 form a connected component.

The algorithm is more confident
about the existence of an edge than about the direction.
Thus the fact that we see cycles 
such as  NR1D1-[ZFP36,ARRDC3]-CD68-NR1D1
is not totally surprising.
What it says is that more information is needed to refine
the relations.  
The causal relation between DDIT3 (aka CHOP) and GDF15 in the context of ER stress is supported by work reported in Li et al \cite{li-zhang-zhong-18bbrc}.
We have not yet found evidence supporting or 
disagreeing with the other hypothesized relations.

The following is a sample of the top 1600 causal relations
proposed by the Time warp causal inference algorithm restricted
to transcripts in Top20X.  
\begin{small}
\begin{verbatim}
ARRDC3 BHLHE40 CPEB4# CREBRF# DDIT3* FAM72D FOS* GADD45A GDF15* 
H1F0 HDAC5# ID4# IER2# ITPKB# NR4A1* PER1 PPP1R15A* SERPINE1 
SLC2A3 TFPI2* THBS2 ZNF550
-b>  [HIST2H2BE*, HSPA5*, RELN]
\end{verbatim}  
\end{small}
\noindent 
Gene set enrichment analysis (GSEA) analysis against the hallmark gene sets in the Broad Molecular
Signatures Database showed overlaps of BHLHE40 FOS DDIT3 SLC2A3
SERPINE1 HSPA5 PPP1R15A with the Hallmark-Hypoxia data set. 
The data set contains 200 genes upregulated in response to low oxygen levels (hypoxia).
BHLHE40 FOS DDIT3 SLC2A3 SERPINE1 HSPA5 THBS2 NR4A1
showed overlaps with the Winter-Hypoxia-Metagene data set. 
The data set contains 242 genes regulated by hypoxia, 
based on literature searches.
In addition, RELN was shown to be regulated by Hif1 alpha and Hif2 alpha in \cite{PMID:14747751}. 
Hypoxia has been known for a long time to cause ER
stress \cite{PMID:16200199} and HIST2H2BE, HSPA5 are among the transcripts identified above as part of the UPR response.
The ENCODE Transcription Factor Binding Site Profiles dataset shows that the promoters for HIST2HBE, HSPA5 and RELN all have binding sites for BHLHE40 and that HIST2HBE and HSPA5 also have binding sites for FOS.


\vspace{-5pt}
\section{Materials and methods}\label{materials-methods} 
\vspace{-5pt}

\subsection{Data Generation}\label{mm-data}

\def\addtm#1{#1\textsuperscript{\tiny TM}}
\def\addrtm#1{#1\textsuperscript{\textregistered}}
\vspace{-5pt}

\paragraph{Cell Culture and Exposure.}

SK-N-AS cells (\addrtm{ATCC} CRL-2137) were seeded at $5\times 10^5$ cells per
well in 12-well plates (Thermo \addtm{Scientific} BioLite 12-556-005) in
Dulbecco’s modified Eagle’s medium (DMEM, \addrtm{ATCC} 30-2002). The
medium was supplemented with $10\%$ fetal bovine serum (FBS,
\addrtm{ATCC} 30-2020), 50 units/mL of penicillin, and 50 µg/mL of
streptomycin (\addtm{Gibco} 15-070-063), and $1\%$ MEM non-essential
amino acids solution (\addtm{Gibco} 11-140-050). Cells were incubated at
$37^\circ$C and $5\%$ CO${_2}$ for 0.5-48 h with or without methamidophos
2 mg/mL (Sigma-Aldrich 33395 and LGC Standards DRE-C14980000). Growth
medium was removed and cells were mixed with RNAprotect Cell Reagent
(Qiagen 76526) and frozen at $-80^\circ$C until analysis.

HepG2/C3A cells (\addrtm{ATCC} CRL-10741) were seeded at 
$1\times 10^5$ cells per well in 12-well plates (Thermo \addtm{Scientific} BioLite 12-556-005) in Eagle’s minimum essential medium (EMEM, \addrtm{ATCC} 30-2003). The medium was supplemented with $10\%$ fetal bovine serum (FBS, \addrtm{ATCC} 30-2020), 50 units/mL of penicillin, and 50 µg/mL of streptomycin (\addtm{Gibco} 15-070-063). Cells were incubated at $37^\circ$C and $5\%$ CO${_2}$ for 1-48 h with or without forskolin 25 µM (R\&D Systems 1099/10).  Growth medium was removed and cells were mixed with RNAprotect Cell Reagent (Qiagen 76526) and frozen at $-80^\circ$C until analysis.

\vspace{-5pt}
\paragraph{RNA Preparation and Microarray Processing.}
Total RNA was extracted from control and treated triplicate exposures using Qiagen's RNeasy Plus Micro kit following the manufacturer's protocol with an additional on-column DNase1 treatment.  RNA concentration (Nanodrop ND-1000, ThermoScientific) and quality (Agilent's 2100 Bioanalyzer with RNA Nano kit) were determined prior to whole transcriptome expression analysis using the \addtm{GeneChip} WT PLUS Reagent Kit and \addtm{GeneChip} Human Transcriptome Array 2.0 microarrays (ThermoFisher Scientific).  According to manufacturer's protocols, 100 ng total RNA was converted to ss-cDNA and 5.5 ${\mu}$g ss-cDNA was fragmented and labeled. Hybridization and scanning were performed by the Stanford University PAN facility. \omitthis{(\url{http://pan.stanford.edu/index.html})}
Data files were processed to determine transcript level average signal intensities using Affymetrix Expression Console (build 1.4.1.46) and Transcriptome Analysis Console 3.0 software and annotation file \url{hta-2-0.na34.hg19.transcript}.


\section{Discussion, Related and Future Work}\label{discussion}

\paragraph{Discussion.}
We confirmed that our suite of analysis algorithms worked 
when applied to data from a new cell line treated with a
different type of challenge without
need to adjust parameters.
We used our novel ranking algorithms combined with significance change
filters to select a set of transcripts as a starting point. Our data
analysis then found four processes as candidate elements of the
broader MoA of methamidophos: UPR, cAMP response, calcium ion related
processes, and cell-cell signaling. For each process we identified
responding transcripts likely to participate in that process using a
combination of GO process annotations, data from other experiments
and pathway databases. We also curated a model of acetylcholine
buildup as a consequence of the inhibition of acetylcholinesterase
(ACHE). The resulting model suggested three downstream effects:
increase in DAG (PMA response), increase in IP3, and active
G-protein-coupled receptor (GPCR). Increase in IP3 induces release of
calcium ion consistent with the identified calcium ion related
response. The cAMP response could be connected to the acetylcholine
signal through the G protein binding partners of ChRm3 or the calcium
ion release.\cite{willoughby-cooper-2007physiolrev} G2 arrest
(identified by cyclin profiles) is a pause in the cell life cycle due
to detected problems, such as DNA damage. Nine of the Top20X
transcripts are annotated with DNA damage related terms. The strong
UPR response suggests another reason for a G2 arrest. We identified a
number of transcripts that are potentially key elements of the MoA
using the GAN fake/real ranking and multirole annotations. Some
participate in already identified processes. The results suggest
investigating cell cycle and metabolic (especially cholesterol)
processes in more detail. Finally, results from two different causal
inference algorithms suggested new connections as well as finding
some known connections.  Although the biological results are maybe
not surprising, we were encouraged by the fact that the pure 
data analysis pulled out transcripts that when further examined
in the light of known biology lead to reasonable hypotheses.

\paragraph{Related work.}

Existing approaches to arrive at MoA candidates from omics data
include enrichment and network perturbation analysis methods to infer
causal networks. Enrichment analysis such as
Ingenuity\textsuperscript{\textregistered} Pathway Analysis
(Qiagen)~~\cite{av-14} and PANTHER~\cite{mi-etal-2017panther} rely on
extensive curated information to identify candidate pathways and
processes. They provide important information but are limited by what
interactions and pathways researcher have focused on and may miss
important aspects. The detecting mechanism of action by network
dysregulation (DeMAND) algorithm \cite{av-16} infers the MoA of small
molecule compounds by looking for dysregulation of their molecular
interactions following compound perturbation. It requires an
interaction network as input as well as gene expression profiles of
control and treated samples. Multiple timepoints can be used, but
they are treated independently. Protein target inference by network
analysis (ProTINA)~\cite{av-17} also uses network perturbation
analysis method for inferring protein targets of compounds from gene
transcriptional profiles. The input network models are specific to
the experimental context and represent transcription factor to gene
interactions. Dynamic models of gene expression are derived from
these networks enabling ProTINA to leverage the information in
timeseries gene expression profiles. Our analysis of the response to
methamidophos used several available knowledge bases to classify
groups of responsive transcripts, but did not require a prior
interaction/gene regulation network.

In \cite{ramsey-etal-2008plos-mphi} the objective is to identify a
set of core differentially expressed transcriptional regulators in
the TLR-stimulated macrophage and the clusters of co-expressed
genes that they may regulate. The methods used include clustering
of expression time profiles; GO annotations of the gene clusters;
time-lagged correlation; and promoter sequence scanning for transcription factor (TF) binding sites recognized by a differentially expressed TF. 
This approach to analysis of timeseries transcriptomics data is
closer in spirit to our work and suggests our analysis might
benefit from looking at promoter sequences to refine the
inferred causal relations.

The analysis was able
to recover many known regulators, and also identified a potential
transcriptional regulator not previously known to play a direct
role in TLR-stimulated macrophage activation, TGIF1. The
review~\cite{zak-aderem-2009imrev-innate} discusses additional
work by the ISB group using similar techniques to infer
transcriptional relations from time series data measured over a
variety of conditions.

\omitthis{
The HPN-DREAM challenge \cite{hill-etal-16dream8} addressed the
problem inferring causal signaling networks in cancer cells.
Phosphoproteomics data was collected for 32 biological contexts (4
cell lines and 8 stimuli; each context perturbed by one of 3
inhibitors plus DMSO control). Expression levels were measured at
11 time points between 5 minutes and 72 hours using reverse-phase
protein lysate arrays with 150-180 antibodies (45 of which were
phospho). The data was split into training and test sets. Teams
participating in the challenge were given training data and
inferred network models were scored on performance on the test
data. The teams used a variety of standard novel methods including
linear/nonlinear regression, bayesian, and ODE based methods with
and without prior knowledge (based on available pathway
databases). No use was made of GO term annotations. A method
called PropheticGranger with heat diffusion prior attained the
best overall score. The small number of proteins measured means
that current machine learning methods are of limited
applicability. Use of inhibitors provides valuable information not
available in pure omics response to a drug or toxin.
}

Li et. al. \cite{li-etal-2012} treated SK-N-SH human neuroblastoma
cells with three organo\-phosphates including methamidophos with high
and low doses for 24h and analyzed the resulting transcriptomics
data. The doses for methamidophos were roughly 10 and 200 times the
concentration used in the present study. The objective was a
comparative study and since they found little common responsed they
combined the data for the three experiments for fold change analysis.
Thus we can not make a meaningful comparison with our results based
on the results reported in the paper.

\paragraph{Future directions.}

Our analysis revealed potentially interesting downstream
effects of methamidophos and  also suggests a number of interesting
future directions.
One is developing methods to better understand which of the multiple
predicted roles different transcripts are playing and when.
Another future direction is understanding the
meaning of edges in graphs synthesized using algorithms with
different assumptions to understand in what sense are they ``causal''.
The two algorithms used in the present study to infer causal relations from timeseries data give
qualitatively different, complementary views of the potential causal
network hidden in the data. The timewarp algorithm produces short
chains (in theMethamidophos case, length 1) and clustered relations--
pairs of groups of compounds where there is an edge from every element
of one group, to every element of the other group as illustrated
above. In contrast, the Siamese Twin algorithm produces longer chains
of causality edges with more diversity of connections (a sparser
graph). One reason is likely the discrete treatment of time and
expression levels used by the timewarp algorithm.  
As both views seem to provide useful information,
more work is needed to understand the precise biological interpretation of the inferred networks.
A first step might be taking into account available promoter binding information, but much more work is needed here.
For example, can we extract interpretable features that the machine learning algorithms find?  
Will that help with understanding the anomalies and the derived ``causal'' relations?
\footnote{Disclaimer.
Research was sponsored by the U.S. Army Research Office and the Defense Advanced Research Projects Agency and was accomplished under Cooperative Agreement Number W911NF-14-2-0020. The views and conclusions contained in this document are those of the authors and should not be interpreted as representing the official policies, either expressed or implied, of the Army Research Office, DARPA, or the U.S. Government. The U.S. Government is authorized to reproduce and distribute reprints for Government purposes notwithstanding any copyright notation hereon.}

\clearpage

\bibliographystyle{splncs03}
\bibliography{methamidophos}


\newpage
\section{Appendix}\label{appendix}
\subsection{Predicate details}\label{predicates}

We define a number of predicates used to select subsets of transcripts
and compare them.   Roughly they concern qualitative measures of
significance (signal vs. noise), measures of change, and ranking functions indicating the presence of distinctive/discriminating features.

\subsubsection{Measures of significance and change} 
\label{significance}

\paragraph{Measures of significance.}
The predicate $\sep(d,n)$ holds for a transcript if 
there is white space (non-overlap/separation) between $d$ standard deviation bands around treated and control GP time profiles for at least $n$ (of $100$) time points.  Thus $\sep(1,15)$ holds for
a transcript if the 1 standard deviation bands around the
GP treated and control time profiles do not overlap for at least
$15$ time points.
A measured time is called \emph{o-significant} if the treated and control measurements define non-overlapping intervals.

\paragraph{Measures of change.}
In the following we give the predicate name
(a short name usable to label rows and columns)
and the definition for measure of change predicates.
\begin{itemize}
\item
2sd : $\sep(2,1)$
\item
1sd : $\sep(1,1)$
\item
1sd33 : $\sep (1,33)$
\item
1sd15 : $\sep(1,15)$
\item
67sd33 : $\sep(.67,33)$
\item
udby1 : change of at least 1 log2 fold (at some o-significant measured time)
\item
udby2 :  change of at least 2 log2 fold (at some o-significant measured time)
\item
nsig6 : at least 6 \emph{o-significant} measured time points
\item
good : udby1 or 2sd
\end{itemize}

\subsubsection{Ranking functions}\label{ranking}

We defined three groups of functions to rank transcripts according
to some measure of possible importance.  The functions are represented
as lists of elements in order of decreasing rank.
\paragraph{PCA.}
There are 6 PCA ranking functions two
for each of the top 3 PCA components (one positive and one negative).
The rank corresponds to the importance/weight of transcript in the given 
PCA component.
\paragraph{Anom.} 
This ranks transcripts according to quality of the
      GP profile reconstructed from the autoencoders representation.
 The worse the reconstruction quality the higher the rank, viewing the
 profile as representing anomalous behavior.\cite{vertes-etal-2018cmsb} 
\paragraph{GAN.}
 Generative Adversarial Network models 
    rank transcripts according to how confident the trained discriminator is that the GP profile represents a transcript.
High fake ranking cooresponds to low confidence that a profile
comes from a transcript (i.e. high confidence
it is not), and high real ranking corresponds to high confidence
that the profile represents a transcript.  
We use three models parameterized by the feature window size 
(Section~\ref{nn-gan}).

\subsubsection{Discriminators.}\label{discriminators}

In the following we give the predicate name
(a short name usable to label rows and columns)
and the definition for discriminator predicates.
\begin{itemize}
\item
t20PCA : ranked in the top 20 in some PCA component
\item
t20an : ranked in the top 20 by some anomaly measure
\item
t20gf : ranked top 20 fake by some GAN discriminator
\item
t20gr : ranked top 20 real by some GAN discriminator
\item
t20 : top20PCA or top20Anom or top20gan-fake or top20gan-real
\item
t20x : t20 and (1sd15 or udby1) also known as Top20X
\end{itemize}

\begin{figure}[h]
\centering
\includegraphics[width=0.8\textwidth]{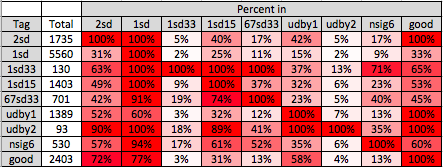}
\hbox{}
\vspace{2mm}
\includegraphics[width=0.8\textwidth]{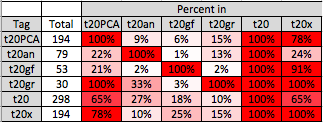}
\hbox{}
\vspace{2mm}
\includegraphics[width=0.8\textwidth]{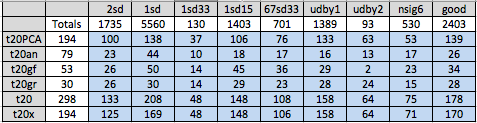}
\caption{Comparing transcript sets}
\label{flc-fld}
\end{figure}

\noindent
In the tables of figure \ref{flc-fld} the size of the different
sets and their intersections are shown.   
In the top two tables the entries $n$ (other than the total column)
is read  $n\%$ of elements in the set given by the row label
belong to the set given by the column label.   In the bottom table
the entries simply represent the size of the intersection.

\subsection{Panther over representation analysis results}
\label{panther}
We used the gene-list analysis service provided by the PantherDB website
(\url{www.pantherdb.org}).  HGNC gene ids were uploaded
from files.  We selected Statistical overrepresentation test,
using the default settings and Homo Sapiens organism.
The results are presented as a tree of GO-slim process terms.
Each node is associated with information includeing the elements of the uploaded gene-list that are annotated with the nodes process term and the fold enrichment.  Results were limited to those with FDR P-value $< 0.05$.

Nine different transcript sets defined using predicates 
of section \ref{predicates} were analyzed.  The following summarizes
the results: 
the predicate, the size of the set (in parens), number of
GOSlim leaves.  Each header is followed by a list of
process terms, each with the number of associated transcripts and 
the fold enrichment.

\begin{itemize}
\item
Top20X (194)  no significant results
\item
udby1 (1389)       no significant results
\item
udby2 (93) 
  \begin{itemize}
  \item
    circadian rhythm  [3, 36]
  \item
    regulation of metabolic  process	[18, 3]
  \end{itemize}
\item
67sd33 (701)
  \begin{itemize}
  \item
  rRNA processing   [15, 4.26]	
    \begin{itemize}
    \item
      RNA metabolic process  [37, 2.3]
      \begin{itemize}
      \item
          aromatic/nucleobase-containing compound metabolic process 
          [52, 2.3]
      \end{itemize}
    \end{itemize}
  \end{itemize}
\item
up log$_2$ fold by 1 hr  (545)  
  \begin{itemize}
  \item
  lymphocyte proliferation	(Interferons) [7, 7]    
  \item
  detection of chemical stimulus involved in sensory 
              perception	(olfactory receptors) [24, 4.5]  
  \item
  localization [23, .48] 
  \item
  cellular metabolic process	[12, .31] 
  \end{itemize}
\item
up log$_2$ fold by 32h (155)  leaves grouped in informal categories
  \begin{itemize}
  \item
  Metabolic (Aldo-keto reductases, R:bile acid synthesis)
  \begin{itemize}
  \item
    prostaglandin metabolic process [3, 57]
  \item
    secondary metabolic process     [3, 28]
  \item
    hormone metabolic process       [4, 23]
   \end{itemize}
  \item
  Growth factor response (negative)
    \begin{itemize}
    \item
    response to fibroblast growth factor  [3, 57]
    \item
    cellular response to growth factor stimulus  [3, 15]
    \end{itemize}
  \item
  negative regulation of MAP kinase activity  [5, 50]
  \item
    (regulation of) circadian rhythm  [3+5 49,35]
  \item
  Chromatin related (6 terms) 
    \begin{itemize}
    \item
     chromatin assembly or disassembly   [3, 30]
    \item
     negative regulation of chromatin silencing    [3, 30]
    \item
     negative regulation of DNA recombination   [3, 21]
    \item
     nucleosome organization     [3, 20]
    \item
     chromosome condensation     [3, 18]
    \item
     chromatin silencing         [4, 14]
   \end{itemize}
  \item
  cellular response to peptide hormone stimulus   [3, 30]
  \item
  response to radiation    [3, 17]
  \item
  cell development    [6, 7]
  \item
  regulation of cell cycle   [9, 7]
  \item
  localization   [3, .21] 
  \end{itemize}

  \item
 dn log$_2$ fold by 32h (146)
  \begin{itemize}
  \item
  mRNA polyadenylation   [4, 21]
  \item
  DNA metabolic process  [12, 5]
  \end{itemize}
\item
up log$_2$ fold by 48h (62)
  \begin{itemize}
  \item   Heat shock related (3 terms total 6 47-92)
  \begin{itemize}
  \item
    cellular response to unfolded protein [3, 92] 
  \item
    cellular response to heat  [4, 47]
  \item
    chaperone-mediated protein folding   [6, 55]
  \end{itemize}
  \item
  nucleosome assembly  [5, 47]  (Histones)
  \end{itemize}

\item
dn log$_2$ fold by 48h (433)
  \begin{itemize}
  \item
  transcription, DNA-templated  [47, 2] (17 ZNFs)
  \end{itemize}
\end{itemize}

\noindent
Notice that localization is underrepresented in
the \emph{67sd33} and \emph{upby 32 h} transcript sets.

\subsection{Cell-cell signaling.}\label{cell-cell}
Following up on the suggestion of cell-cell signaling response
from the Miru analysis, we found that 
three $k$-means clusters have cell-cell signaling annotation.
Interestingly, these clusters also have the G protein-coupled receptor
signaling pathway annotation. Members of these clusters all have an
upregulation peak at 1 h.
No members of theses clusters with cell cell signaling annotation are in Top20X, so we
selected transcripts that are either up or downregulated by 1 log$_2$ fold at some time, or pass the 2 standard deviation separation test.
We call these `good'.   
\begin{itemize}
\item
Cluster 55 has 7 good transcripts with cell-cell signaling annotation:
    CCR5 FGF16 FGF6 GJA3 IAPP IFNA2 IL17A
\item
Cluster 116 has 2 good transcripts with cell-cell signaling annotation:    CCL13	IFNA7
\item
Cluster 124 has 1 good transcript with cell-cell signaling annotation:  CCL7 
\end{itemize}
Figure \ref{cell-cell} shows the time profiles of these transcripts.
We see the 1 h spike followed by rather unsettled but low level
response.

\begin{figure}[h]
\centering
\includegraphics[width=0.8\textwidth]{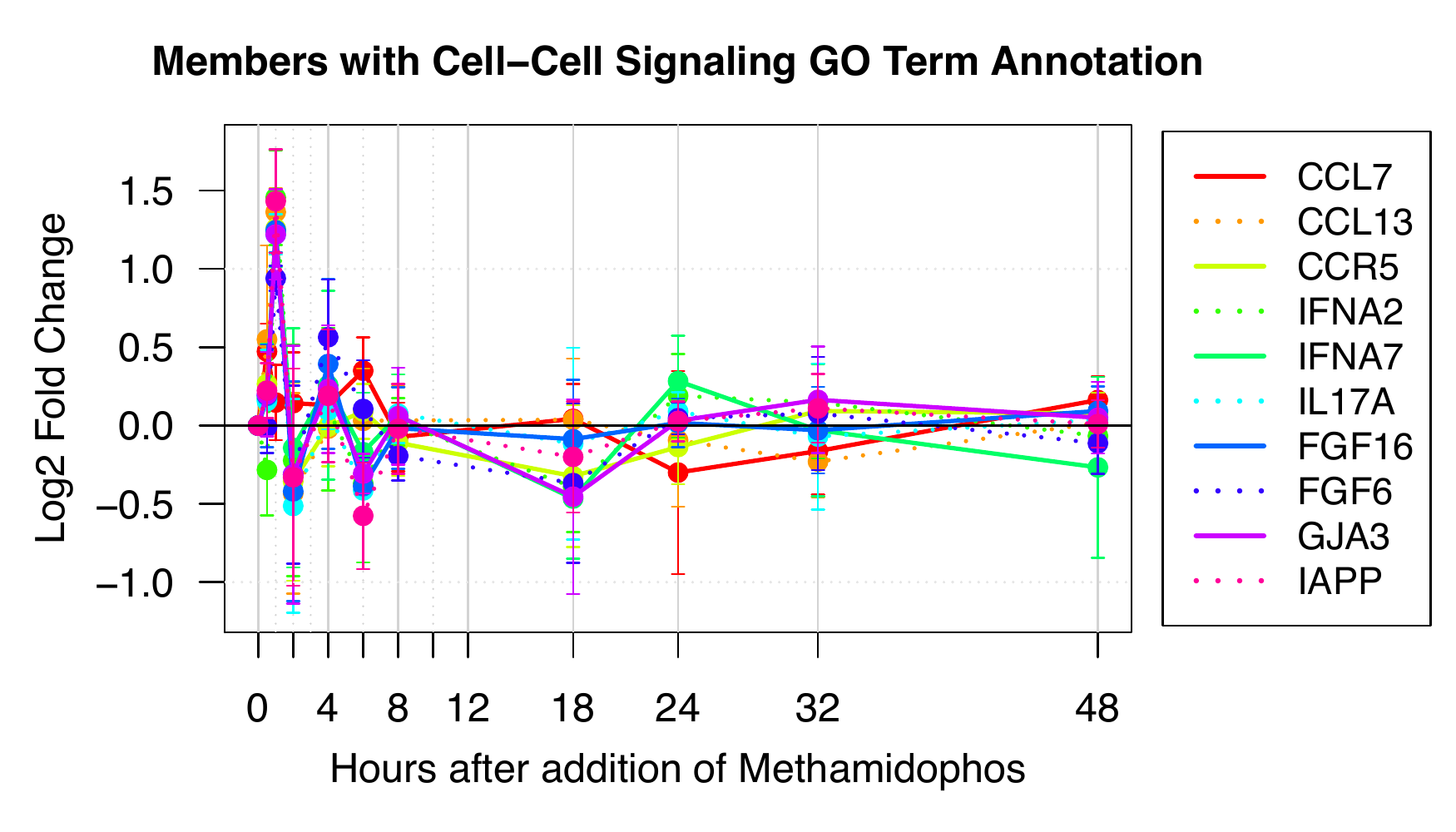}
\caption{Basic time profiles of cell-cell signaling annotated transcripts}
\label{cell-cell}
\end{figure}

\noindent
FGF6, FGF16 are growth factors and GJA3 is a gap junction protein. 
According to Wikipedia, IAPP (aka DAP, Amylin, or islet amyloid polypeptide), is a 37-residue peptide hormone that selectively regulates glucose metabolism.  The rest are immune response proteins.

We checked the Top20X transcripts and found that only one,
GDF15, is annotated with cell-cell signaling.  The basic time
profile of GDF15 has .7 log$_2$ ratio at 1 h, but then no 
measurable response until 32 h (where its has a 2 log$_2$ ratio).
Clearly, having data measuring what is secreted would help to
determine if cell-cell signaling is a real response and
how it fits with other processes.

\omitthis{
GDF15 annotated with cell-cell signaling. Is it secreted?
(apply prDarr (invoke lfmap-u6 "get" (apply hs2tcf-u6 "GDF15")))
0.0659 0.7412 -0.1439 0.0384 -0.0047 -0.0331 -0.1565 0.2055 2.0718 3.707
(apply prArr (invoke sigmap-u6 "get" (apply hs2tcf-u6 "GDF15")))
[false, true, false, false, false, false, false, false, true, true]
}

\subsection{Distinguished transcripts}\label{supp-distinguish}

\paragraph{GAN fake cluster details.}
In the following we list each GAN fake cluster, its
GAN fake members and key associated GO terms, if any.
I the first item ``kmr27'' identifies $k$-means ratio cluster 27.
``(+32:*)'' says the cluster is upregulated (+) at 32h and stays
up (*) (to the end of the measured times).
``9 fake of 10 Top20X of 24 total'' says the cluster has 24 members,
10 of which are Top20X, and 9 of these are ranked GAN fake.
\begin{itemize}
\item
kmr27 (+32:*) 9 fake of 10 Top20X of 24 total
\begin{itemize}
\item
CREBRF - negative regulation of endoplasmic reticulum unfolded protein response
\item
DIAPH2 - actin filament organization
\item
ERMP1 - metal ion binding
\item
HDAC5 - chromatin organization; response to LPS and to insulin
\item
IDS - glycosaminoglycan catabolic process
\item
LACTB - regulation of lipid metabolic process
\item
RASA4 - Ca++; negative regulation of Ras protein signal transduction
\item
SPRY4 - negative regulation of Ras protein signal transduction
\item
ZNF550 
\end{itemize}
\item
kmr78 (+32:*) 5 fake of 10 Top20X of 39 total
\begin{itemize}
\item
AHNAK - regulation of voltage-gated calcium channel activity
\item
ARMCX1 
\item
ITPKB -  Ca++; inositol phosphate biosynthetic process
\item
SMAD9 - cellular response to BMP stimulus
\item
TBX20 - negative regulation of SMAD protein complex assembly; muscle contraction
\end{itemize}

\item 
kmr127 (+32:*) 4 fake of 6 Top20X of 28 total
\begin{itemize}
\item
GJA1 - cell communication by electrical coupling; positive regulation of I-kappaB kinase/NF-kappaB signaling
\item
ID4 - transcription factor;  circadian rhythm
\item
IER2 - cell motility
\item
MIDN - negative regulation of glucokinase activity;
\end{itemize}
     
\item
kmr81  (min@8h) 7 fake of 7 Top20X of 36 total  
\begin{itemize}
\item
ARFGAP2 - endoplasmic reticulum to Golgi vesicle-mediated transport; GTPase activator activity
\item
DHCR7 - cholesterol biosynthetic process; 7-dehydrocholesterol reductase activity
\item
HMGCS1 - cellular response to cholesterol; cholesterol biosynthetic process
\item
METTL12 - protein methylation
\item
NSDHL - cholesterol biosynthetic process
\item
TSPYL4 
\item
UBXN8 - ubiquitin-dependent ERAD pathway (targeting ER-resident proteins for degradation)
\end{itemize}
   
\item
kmr20 (-48:*) 4 fake of 4 Top20X of 58 total
\begin{itemize}
\item
ACAT2 - cholesterol esterification; cholesterol O-acyltransferase activity; cholesterol biosynthetic process
\item
LANCL2 - negative regulation of transcription
\item
PHF23 - positive regulation of protein ubiquitination
\item
TOE1 - (Target of EGR1) RNA phosphodiester bond hydrolysis,
\end{itemize}

\item
kmr38 (-48:*) 6 fake of 10 Top20X of 46 total
\begin{itemize}
\item
ARHGAP28 - negative regulation of stress fiber assembly
\item
FNTB - protein farnesylation;
\item
GNG2 - G protein-coupled receptor signaling pathway; adenylate cyclase-activating 
\item
IGFBP4 - regulation of glucose metabolic process
\item
NEBL - actin filament binding
\item
PIGA - phosphatidylinositol N-acetylglucosaminyltransferase 
\end{itemize}
\end{itemize}
\noindent
Note that Cluster 81 has several cholesteral related transcripts.
Also (not shown) the graphs generated by several algorithms
predict that HMGCS1 is well-connected.  It is interesting to
consider if some Cholestorol pathway has a specific role or
it is a generic response.

\paragraph{GAN real clusters.}
In the following we list each GAN real cluster, its
GAN real members and key associated GO terms, if any.
The notation is the same as for GAN fake clusters.
\begin{itemize}
\item
kmr83 (+1:*) 1 real of 1 Top20X of 1 total
\begin{itemize}
\item  TIPARP - Acts as a negative regulator of AHR which inhibits PER1 expression cellular response to organic cyclic compound 
\end{itemize}

\item 
kmr44 (+2:*)  1 real of 1 Top20X of 1 total
\begin{itemize}
\item CYP1B1 - toxin metabolic process 
\end{itemize}

\item
kmr10 (+18-32:*), 3 real of 3 Top20X of 3 total
\begin{itemize}
\item
NR4A1 - cAMP; negative regulation of cell cycle 
\item
VGF - cAMP; cellular protein metabolic process; glucose homeostasis; insulin secretion;
\item
ZFP36 - cAMP; mRNA catabolic process; negative regulation of polynucleotide adenylyltransferase activity;
\end{itemize}

\item
kmr42 (+18:*)  1 real of 1 Top20X of 1 total
\begin{itemize}
\item  EGR1 - transcription factor; protein refolding; negative regulation of cell proliferation; regulation of protein ubiquitination
\end{itemize}

\item
kmr112 (+32:*)  4 real of 4 Top20X of 4 total
\begin{itemize}
\item
DDIT3 - UPR; Ca++; cell cycle arrest; cellular response to DNA damage stimulus
\item
GDF15 - cAMP; BMP signaling pathway; cell-cell signaling;
\item PPP1R15A - UPR; cell cycle arrest; cellular response to DNA damage stimulus; negative regulation of protein dephosphorylation 
\item TFPI2 - extracellular matrix structural constituent; 
serine-type endopeptidase inhibitor activity
\end{itemize}

\item
kmr63  (-2,4:*)  2 real of 2 Top20X of 2 total
\begin{itemize}
\item
RHOB -  actin filament bundle assembly; positive regulation of angiogenesis;  endosome to lysosome transport
\item
TXNIP - Ca++; cell cycle; negative regulation of cell division; 
response to hydrogen peroxide/glucose
\end{itemize}

\item
kmr85  (-18:*)  1 real of 1 Top20X of 1 total
\begin{itemize}
\item NEUROG2 - E-box binding transcription factor
\end{itemize}

\item
kmr71 (-18:*)  3 real of 3 Top20X of 3 total
\begin{itemize}
\item
HAND1 - transcription factor
\item
RGS4 - negative regulation of G-protein coupled receptor protein signaling pathway; G-protein alpha-subunit binding calmodulin binding;
\item
POPDC2  - cAMP-binding, regulation of membrane potential
\end{itemize}

\item
kmr87 (-18,24:*)  2 real of 2 Top20X of 2 total
\begin{itemize}
\item
CACNG4 - membrane depolarization; neurotransmitter receptor internalization; positive regulation of AMPA receptor activity 
\item
NCOA5 -  glucose homeostasis; negative regulation of insulin receptor signaling pathway 
\end{itemize}
\end{itemize}

\end{document}